\def\mearth{{\rm\,M_{Earth}}}
\def\rearth{{\rm\,R_{Earth}}}
\def\msun{{\rm\,M_{Sun}}}
\def\rsun{{\rm\,R_{Sun}}}
\def\lsun{{\rm\,L_{Sun}}}
\def\gsim{~\rlap{$>$}{\lower 1.0ex\hbox{$\sim$}}}
\def\lsim{~\rlap{$<$}{\lower 1.0ex\hbox{$\sim$}}}

\def\wpmsq{W m$^{-2}$}

\def\eg{{\it e.g.\ }}

\def\ie{{\it i.e.\ }}
\def\cf{{\it c.f.\ }}

\documentclass[12pt]{article}
\usepackage{graphicx}
\usepackage{appendix,amsmath}
\usepackage[round]{natbib}
\begin{document}

\begin{center}\Large{Tidal Venuses: Triggering a Climate Catastrophe via Tidal Heating}\\
\smallskip\normalsize
Rory Barnes$^{1,2,3}$, Kristina Mullins$^{1,2}$, Colin Goldblatt$^{1,2,4}$, Victoria S. Meadows$^{1,2}$, James F. Kasting$^{2,5}$, Ren{\'e} Heller$^6$
\end{center}

\begin{abstract}
Traditionally stellar radiation has been the only heat source
considered capable of determining global climate on long
timescales. Here we show that terrestrial exoplanets orbiting low-mass
stars may be tidally heated at high enough levels to induce a runaway
greenhouse for a long enough duration for all the hydrogen to
escape. Without hydrogen, the planet no longer has water and cannot
support life. We call these planets ``Tidal Venuses,'' and the
phenomenon a ``tidal greenhouse.''  Tidal effects also circularize the
orbit, which decreases tidal heating.  Hence, some planets may form
with large eccentricity, with its accompanying large tidal heating,
and lose their water, but eventually settle into nearly circular
orbits (i.e.  with negligible tidal heating) in the habitable zone
(HZ). However, these planets are not habitable as past tidal heating
desiccated them, and hence should not be ranked highly for detailed
follow-up observations aimed at detecting biosignatures. We simulate
the evolution of hypothetical planetary systems in a quasi-continuous
parameter distribution and find that we can constrain the history of
the system by statistical arguments. Planets orbiting stars with
masses $\lsim 0.3~\msun$ may be in danger of desiccation via tidal
heating. We apply these concepts to Gl 667C c, a $\sim 4.5~\mearth$
planet orbiting a $0.3~\msun$ star at 0.12~AU. We find that it
probably did not lose its water via tidal heating as orbital stability
is unlikely for the high eccentricities required for the tidal
greenhouse. As the inner edge of the HZ is defined by the onset of a
runaway or moist greenhouse powered by radiation, our results
represent a fundamental revision to the HZ for non-circular orbits. In
the appendices we review a) the moist and runaway greenhouses, b)
hydrogen escape, c) stellar mass-radius and mass-luminosity
relations, d) terrestrial planet mass-radius relations, and e) linear
tidal theories.

\end{abstract}

\medskip

\noindent$^1$Astronomy Department, University of Washington, Box 951580, Seattle, WA 98195\\
$^2$NASA Astrobiology Institute -- Virtual Planetary Laboratory Lead Team, USA\\
$^3$E-mail: rory@astro.washington.edu\\
$^4$Department of Earth and Ocean Science, University of Victoria, Victoria, BC\\
$^5$Department of Geosciences, The Pennsylvania State University, State College, PA\\
$^6$Leibniz Institute for Astrophysics Potsdam (AIP), An der Sternwarte 16, 14482 Potsdam, Germany

\clearpage

\section{Introduction \label{sec:intro}}

All life on Earth requires liquid water. Not surprisingly, our search
for life on exoplanets therefore begins with identifying environments
that support it \citep{Hart79,Kasting93,Selsis07,vonBloh07}. Many other
features of a planet are also important, such as the mix of gases in
the atmosphere and the interior's structure and energy budget, but current research has focused primarily on
the stability of surface water. On Earth, liquid water persists
primarily because the Sun's radiation heats the surface to a
temperature between water's freezing and boiling points. Thus, the
concept of a ``habitable zone'' (HZ) emerged, which is the region
around a star in which insolation can maintain liquid water on the
surface, assuming an Earth-like planet.

Stars have a wide range of luminosities, most of which are
considerably lower than the Sun's, yielding HZs that are closer
in. At the extreme, the luminosity becomes so low that a planet
orbiting at the stellar surface would be too cold to support liquid
water on its surface. Such a star has no
HZ. The recently-discovered Y dwarf (which is not a star) WISEP J1828+2650
\citep{Cushing11}, with an effective temperature below 300 K, is an
example of such a primary. For warmer stars, HZs may still be close
enough in that non-radiative processes may impact habitability, such
as stellar flaring
\citep[\eg][]{Lammer07,Khodachenko07,Tian09,Segura10,Lammer10},
decreased initial volatile inventory \citep{Raymond07,Lissauer07} or tidal
effects
\citep[\eg][]{Kasting93,Joshi97,Jackson08_hab,Correia08,Barnes09_THZ,Heller11}. As the HZ
of our Sun is too distant for these phenomena to affect the Earth, we
can currently only explore their role theoretically, and, consequently, many
scientists consider close-in planets less favorable candidates for
habitability.

Nevertheless, the last few years have seen renewed interest in the
potential habitability of planets in orbit about low luminosity
objects
\citep{Tarter07,Scalo07,Lunine08,Monteiro10,Agol11,Bolmont11}. This shift occurred because terrestrial-sized planets are easiest to
detect around low luminosity hosts, due to the larger mass and radius
of the planet relative to the star. Furthermore, these objects are the
most abundant in the solar neighborhood. Planets with masses between 1
and 
$10~\mearth$, ``super-Earths,'' have indeed been detected in the last
few years around low-mass stars, such as Gl~581 d and Gl~667C c by
radial velocity
\citep{Udry07,Mayor09,Vogt10,Forveille11,Bonfils11,AngladaEscude12,Delfosse12}
and GJ~1214 b by transit
\citep{Charbonneau09,Sada10,Kundurthy11,Carter11}. Several observational campaigns designed specifically to detect planets around low mass stars
are now underway
\citep{NutzmanCharbonneau08,Boss09,Zechmeister09,Bean10,Rodler11}.

The inner edge of the HZ is especially prone to non-radiative
phenomena. Traditionally, the inner boundary is defined by the stellar
distance at which insolation is strong enough to remove all water in the atmosphere and on the surface
\citep{Kasting88}. Two different dehydrating scenarios (the ``moist'' and ``runaway'' greenhouse) are discussed in detail in
$\S$~\ref{sec:innerIHZ} and App.~\ref{app:dg}. Both require water
vapor to penetrate a stratosphere, be dissociated (photolyzed) by
high energy radiation, and culminate in the escape of hydrogen. We call any
process that can ultimately lead to total water loss a ``desiccating
greenhouse.'' Without sufficient hydrogen, water cannot form, and the planet
will remain uninhabitable forever, unless a major, unlikely event
occurs, \eg an impact by a water-rich body that simultaneously delivers water and changes the atmosphere in a way that halts the desiccating greenhouse. This definition of the inner edge is conservative because no
known process can maintain habitability against a desiccating
greenhouse. Other processes could be equally deleterious for life, but
clearly total desiccation will terminate habitability.

We will discuss four types of terrestrial planets in this study,
classified by their water content. ``Wet'' planets are terrestrial
exoplanets that have a water content similar to the Earth. ``Dry''
exoplanets have far less, $\sim 3$ cm deep if condensed and spread
globally, but are habitable \cite[see][]{Abe11}. ``Desiccated''
planets have lost all their water through a desiccating
greenhouse. Finally, ``water worlds,''
\citep{Raymond04,Leger04} are planets with a much larger
inventory of water than the modern Earth, \eg a warm Europa. This investigation explores
the transition of wet, dry and water worlds into the desiccated state.

\cite{Dole64} was the first to point out that terrestrial planets
in the HZ of low luminosity stars can have their spin altered by tidal
interaction. In particular, the danger of synchronous rotation, \ie
one hemisphere always facing the star, was
emphasized. \cite{Kasting93} quantified this concept and found that
planets orbiting within the HZ of stars less than two-thirds the mass of the Sun were
in danger of synchronization. Although their analysis was limited to
Earth-like planets on circular orbits, a general belief developed that
those planets could not be habitable, as one half of the planet would freeze while the other would roast.

Synchronization is certainly an important consideration when assessing
habitability, and many investigations have explored its role, but with
mixed results. Atmospheric modeling initially suggested that
circulation will transport energy to the unlit side, ameliorating the
extreme temperature difference \citep{Joshi97}.  Some subsequent
modeling has confirmed that synchronous rotators are likely to have
super-rotating atmospheres
\citep{Joshi03,HengVogt11,Edson11,ShowmanPolvani11}, while others have
discounted it \citep{Wordsworth11}, and still others have suggested
that such a state might be beneficial at the outer edge of the HZ
\citep{Pierrehumbert11}. Taken together, these investigations suggest
that synchronized planets should not be dismissed uniformly as
uninhabitable. Hence, synchronization is not as stringent a constraint as
desiccation, and therefore is not an HZ boundary.

For many years, confusion also existed regarding the term ``tidal
locking.'' Many investigators assumed it was synonymous with
``synchronous rotation.'' If the orbits are non-circular, as for many exoplanets
\citep{Butler06}, then tidally-evolved planets may reach an equilibrium state where they rotate faster than
synchronous with an ``equilibrium'' or ``pseudo-synchronous''
period. This aspect of tidal theory has been known for decades
\citep[\eg][]{Goldreich66,GreenbergWeidenschilling84}, but has only recently been pointed out for the
case of exoplanets
\citep{Barnes08_hab,FerrazMello08,Correia08}. Therefore, some
exoplanets, such as Gl 581 d with an eccentricity of 0.38
\citep{Mayor09}, may be ``tidally locked'' but rotate about twice per orbit
\citep{Barnes08_hab,Heller11}. For more on this point, consult
\citet{MurrayDermott99}, chap.~5.2, and for a counterargument see
\cite{MakarovEfroimsky12}. In this paper, we use ``tidally locked'' to mean a planet rotating at the equilibrium period as determined by its eccentricity and obliquity, see Eqs.~(\ref{eq:p_eq_cpl}), (\ref{eq:p_eq_ctl_obl})--(\ref{eq:p_eq_ctl}). In summary, a synchronously rotating planet is tidally locked (yet, it
could still have non-zero eccentricity and obliquity), but a tidally
locked planet is not necessarily rotating synchronously.

Even if an orbit is currently circular, tides may not have driven the
rotation to synchronous. If the orbit began with large eccentricity,
tides will tend to damp it to zero and we may expect it to be rotating
synchronously. However, the planet could pass through one or more
``spin-orbit resonances,'' where the planet's rotational frequency is
commensurate with its orbital frequency \citep[see \eg][]{Rodriguez12}. For example, Mercury rotates
three times for every two times it orbits the sun, a 3:2 spin-orbit
resonance. Spin-orbit resonances require an inhomogeneous mass
distribution, which is likely for tidally-deformed exoplanets, but
cannot be measured for the foreseeable future. A planet caught in a
spin-orbit resonance may remain in that state even if circularized, as
a resonance is a strong dynamical process. For any particular
exoplanet, capture and retention into a spin-orbit resonance will be
very difficult to constrain observationally, so all reasonable options
should be considered.  Therefore, synchronous rotation is unlikely for
planets with large eccentricities (Mercury's is 0.2), and not even
guaranteed for a circular orbit. For more on spin-orbit resonances,
the reader is referred to \citet{MurrayDermott99}, chap.~5.4.

As tidal locking of the planetary rotation is not an absolute
constraint on habitability, we turn to tidal heating as the other
tidal phenomenon most likely to affect planetary habitability. As a
planet moves from periastron, its closest approach to the star, to
apoastron, the furthest point, and back again, the gravitational force
changes, being inversely proportional to distance squared. This
difference creates an oscillating strain on the planet that causes its 
shape to vary periodically. The rigidity of the planet resists
the changes in shape, and friction generates heat. This energy production
is called tidal heating.

Tidal heating is responsible for the volcanism on Io
\citep{Strom79,Laver07}, which was predicted, using tidal theory, by \citet{Peale79}. Io is a small body orbiting
Jupiter with an eccentricity of 0.0041, which is maintained by the
gravitational perturbations of its fellow Galilean moons, that shows
global volcanism which resurfaces the planet on a timescale of 100 --
$10^5$ years
\citep{Johnson79,Blaney95,McEwen04}. The masses of Jupiter and Io are orders
of magnitude smaller than a star and terrestrial exoplanet, and thus
the latter have a much larger reservoir of orbital and rotational energy
available for tidal heating. Moreover, some exoplanets have been found
with orbital eccentricities larger than 0.9
\citep{Naef01,Jones06,Tamuz08}. Thus, the tidal heating of terrestrial
exoplanets may be much more effective than on Io
\citep{Jackson08_heat,Jackson08_hab,Barnes09_THZ,Barnes10_corot7,Heller11}. This expectation led
to the proposition that terrestrial exoplanets with surface heat
fluxes as large or larger than Io's should be classified as
``Super-Ios'', rather than ``Super-Earths''
\citep{Barnes09_40307}. Numerous Super-Io candidates exist, such as CoRoT-7 b \citep{Leger09,Barnes10_corot7}, Gl 581 e \citep{Mayor09}, 55 Cnc e \citep{McArthur04,DawsonFabrycky10,Winn11}, and Kepler-10 b \citep{Batalha11}, but none is in the HZ.

\citet{Jackson08_hab} and \citet{Barnes09_THZ} considered Io's heat flux,
$\sim 2$~\wpmsq~\citep{Veeder94,Spencer00,McEwen04}, to be an upper
limit for habitability, arguing that Io-like surfaces are dangerous
for habitability. However, 2~\wpmsq~may not be sufficient to sterilize a
planet, and so should not be considered a hard limit to
habitability. For example, the heat flow in inhabited hydrothermal
vent systems on Earth, such as the Endeavour segment of the Juan de
Fuca Ridge \citep{Holden98}, is $\sim 30$~\wpmsq~
\citep{Fontaine11}. Thus, a water world with tens of
\wpmsq~of energy output could support life. While Io-like
volcanism is clearly an issue for habitability, it may not always lead
to sterilization. However, if the tidal heating can maintain a
desiccating greenhouse long enough for
all the water to be lost, then the planet becomes uninhabitable, and is
highly unlikely to ever regain habitability.

Calculations of the inner edge of the HZ have traditionally assumed
the primary energy source at the surface is stellar radiation, as is
the case for the Earth. Following \citet{Barnes09_THZ}, we call that
type of HZ an ``insolation HZ'' (IHZ), as starlight is the only energy
source considered. In this investigation, we identify
the amount of tidal heating that triggers a desiccating greenhouse, as
well as the combinations of physical and orbital parameters for which
tidal heating could yield it. We find such a state is predicted by
current models, and dub such a world a ``Tidal Venus,'' and the
phenomenon that produces it a ``tidal greenhouse.'' The tidal
greenhouse would probably have the same effect on habitability as one
caused by irradiation, and hence should be considered as hard a limit
to habitability as a traditional, insolation-driven desiccating
greenhouse.

In this study, we define the limits of Tidal Venuses in terms of
stellar and planetary mass, $M_*$ and $M_p$, respectively, the orbital
semi-major axis $a$, orbital eccentricity $e$, planetary radius $R_p$,
planetary obliquity $\psi_p$, and planetary spin frequency
$\omega_p$. Some of these quantities are easily observed by current
technology, others require special geometries and the next generation
of space telescopes. Therefore, for the next decade, not all
newly-found planets in an IHZ will have well-constrained tidal
heating. This study provides a framework for identifying the range of
tidal heating on terrestrial planets orbiting in the IHZ of low
luminosity stars. As we see below, the story is complicated and
involves a large parameter space, but is tractable.

We first review the surface conditions that lead to a greenhouse state
in $\S$~\ref{sec:innerIHZ}, including an extended discussion of the
moist and runaway greenhouses (App.~\ref{app:dg}), a short
review of hydrogen escape driven by high energy radiation
(App.~\ref{app:atmloss}), the relationships between mass, radius and
luminosity for low-mass, hydrogen burning stars, ``M dwarfs'' or
``late-type stars'' in the parlance of astrophysics, the extent of
the IHZ (App.~\ref{app:relations}), and mass-radius relationships for
terrestrial exoplanets (App.~\ref{app:terrmr}). We then briefly
describe tidal heating in $\S$~\ref{sec:tides} with details in
App.~\ref{app:tides}. Next we show that tidal heating alone can
produce surface conditions capable of triggering a runaway greenhouse
on planets orbiting M dwarfs ($\S$~\ref{sec:tv}). Then we explore how
past tidal heating may preclude habitability of planets found in the
IHZ regardless of their current tidal heating
($\S$~\ref{sec:history}). In $\S$~\ref{sec:gl667c} we then consider
the Gl 667C system which contains two potentially habitable planets
and find that tidal heating is unlikely to have sterilized either. In
$\S$~\ref{sec:disc} we discuss the results, and finally, in
$\S$~\ref{sec:concl} we draw our conclusions.

%
%
%
%

\section{The Inner Edge of the Habitable Zone \label{sec:innerIHZ}}


Planets with surface water will always have water vapor in their
atmospheres. The amount of water vapor present in equilibrium with a
liquid surface is described by the saturation vapor pressure $p_{sat}$,
which is the pressure exerted by water vapor in thermodynamic
equilibrium with standing surface water. The value of $p_{sat}$ depends exponentially on temperature: a warmer planet will have
much more water vapor in its atmosphere. Water vapor is a greenhouse
gas (see below), so a warmer planet will have a stronger greenhouse
effect, enhancing warming, a positive (but not necessarily runaway) feedback.

Atmospheric gases on Earth---and those that we expect on habitable
Earth-like planets---are mostly transparent to optical wavelengths, so
starlight is able to heat the surface. The surface in turn heats the
air in contact with it, which rises, expanding and cooling
adiabatically. As the air cools, water vapor condenses,
releasing latent heat and so slowing the cooling. The net rate of
temperature decrease with height is called the ``moist adiabatic lapse
rate,'' setting the mean temperature--pressure
($T$--$p$) structure of Earth's troposphere (the lower region of the
atmosphere affected by convection, and bounded at the bottom and top
by the planetary surface and the tropopause, the altitude at which
temperature stops dropping with height).

Atmospheric gases which absorb radiation at similar wavelengths to the
radiative emission of the planet (thermal infrared for Earth and any
habitable planet) are termed ``greenhouse gases''; these cause a
``greenhouse effect''. The most important greenhouse gases for Earth
are water vapor and carbon dioxide. They absorb radiation emitted by
the surface, and then re-emit, both downward toward the surface and
upward into space. Energy radiated towards the surface (commonly called back
radiation) heats the surface. Most importantly, as the atmosphere is
cooler than the surface, the amount of radiation emitted to space from
the atmosphere is less than the amount of energy emitted by the surface. Thus, the presence of a greenhouse atmosphere means
the surface temperature is warmer than the effective temperature of
the planet.

Earth's greenhouse effect keeps the planet warm enough to be
habitable; without it Earth's surface would be at the planet's
effective temperature, a barren 255\,K. However, stronger and more
water-vapor-rich greenhouse atmospheres can render the planet
uninhabitable through high temperature sterilization and/or
desiccation (loss of the ocean), and thus define the inner edge of the
habitable zone. There are two physically distinct situations that lead to loss of habitability, the ``runaway greenhouse'' and ``moist greenhouse''.

The runaway greenhouse was recently reviewed by
\citet{GoldblattWatson12}, so we provide a summary description
only. As the planet warms, the amount of water vapor in the atmosphere
increases such that water becomes a major constituent of the
atmosphere and ultimately the dominant one. Consequently, the moist
adiabatic lapse rate tends toward the saturation vapor pressure curve for
water and the $T$--$p$ structure of the atmosphere becomes
fixed. Concurrently, the atmosphere becomes optically thick in the
thermal infrared, such that only the upper troposphere can emit to
space. As the $T$--$p$ structure is fixed, the emitted radiation is also
fixed, imposing a limit on the outgoing radiation ($F_{crit}$) from
the troposphere. Values for $F_{crit}$ from the literature are
typically 285\,W\,m$^{-2}$ to 310\,W\,m$^{-2}$ for a $1~\mearth$ planet
\citep{Pollack71,Watson84,AbeMatsui88,Kasting88,Pierrehumbert10}. The
physics of this limit is described in more detail in \citet{Simpson27} and
\citet{Nakajima92}. (\citet{Komabayashi67} and \citet{Ingersoll69} describe a
stratospheric limit at a higher flux, 385\,W\,m$^{-2}$, which is never
reached in practice.) If the amount of energy supplied to the
atmosphere by the Sun \citep{Simpson27}, impacts \citep{AbeMatsui88},
or tidal heating (this work) was to exceed $F_{crit}$ then the
atmosphere would not be able to maintain radiation balance and runaway
heating of the surface would ensue, causing evaporation of the entire
ocean. Radiation balance would be regained when either a) the surface
temperature reaches $\sim 1400$~K at which point enough radiation is
emitted in the near infrared where water vapor is not a good absorber,
or b) if all the water vapor is lost from the atmosphere.

One likely water loss mechanism is hydrogen escape to space. Today,
little water vapor reaches the upper atmosphere because the tropopause
acts as a ``cold trap'': almost all water has condensed lower in the atmosphere, so the tropopause
water vapor mixing ratio (the abundance of water vapor in the atmosphere,
expressed as the ratio of the mass of water vapor to the mass of dry air) is
low. Water
vapor transport above here is generally diffusive, leading to a
constant water vapor mixing ratio in the upper atmosphere. (More
precisely, in Earth's atmosphere, water vapor increases with altitude
in the stratosphere due to methane oxidation and decreases with
altitude in the next higher atmospheric layer, the mesosphere, due to photolysis; however, the total
hydrogen mixing ratio is conserved, and the hydrogen escape rate depends on this value.) Thus,
water is not a strong source of escaping hydrogen and the overall
escape rate is low. However, in the runaway greenhouse the
atmosphere becomes predominantly water, so no such constraint applies and
hydrogen may escape hydrodynamically to space
\citep{KastingPollack83}. The rate will depend on the amount of
extreme ultraviolet (XUV, 1 -- 1200\AA) radiation absorbed in
the highest part of the atmosphere, so this process is also limited by
the available stellar XUV energy
\citep{Watson81}. A Venus-size planet at the inner edge of the
habitable zone, could have lost an ocean the size of Earth's in $\sim
10^{8}$ years \citep{Watson81}; we refer to this as the desiccation
time, $t_{des}$ (discussed further below). Deuterium would be retained
preferentially over ordinary hydrogen during hydrogen
escape. Enrichment of D/H on Venus implies that Venus lost a
substantial amount of water to space \citep{Donahue82,deBergh93}, likely
after experiencing a runaway greenhouse. 

We adopt $t_{des} = 10^8$ years based both on the work on Venus,
as well as calculations presented in App.~\ref{app:atmloss} for a
terrestrial planet orbiting a very low mass star. Such stars are
very active and emit strongly in the near UV, especially when they are
young
\citep[\eg][]{West08}, but the total flux is still less than the
present-day Sun \citep{Fleming93} and the XUV flux (which drives
hydrodynamic escape) is not well known. Thus, we expect a 
range of $t_{des}$ to exist, especially since the orbit evolves with time
due to tides, but choose this value for simplicity. The model described
in App.~\ref{app:atmloss} can be applied to individual cases when
observations permit realistic modeling. We stress that there are many
unknowns which may affect the actual value of $t_{des}$. The masses of
exoplanet oceans are unknown. Earth's mantle contains a few ocean
masses of water
\citep{BellRossman92,Murakami02,MartyYokochi06}, so it is possible
that the surface of a planet could acquire a new ocean via
outgassing after
desiccation, and become habitable again. Thus, we cannot conclude
with certainty that a planet with a desiccated atmosphere will never
be habitable in the future. However we also expect rapid mantle
overturn during strong tidal heating, and hence the entire water
inventory may actually be lost. We therefore assume that the tidal greenhouse is highly likely to
permanently sterilize a planet. 

The moist greenhouse \citep{Kasting88} describes a warm atmosphere, in
which the whole troposphere is assumed to be water vapor saturated and is underlain by a liquid ocean.
As the surface temperature increases, the tropopause is pushed higher
(to lower pressure) given the reasonable assumption of a constant
tropopause temperature. While the saturation vapor pressure at the
tropopause is, by this assumption, constant, the saturation mixing
ratio of water at the tropopause ($p_\text{sat}/p_\text{trop}$)
increases as the tropopause moves higher, where $p_\text{trop}$ is the
atmospheric pressure at the tropopause. The cold
trap is then no longer effective, and substantial water vapor penetrates the
stratosphere, and water-derived hydrogen escape can be effective. The
planet may thus gradually desiccate while in a hot, but stable, climate. This
process could be driven by high greenhouse gas inventories rather than
external heating.

A key distinction between the runaway and moist greenhouses is that
the former happens when a known flux of energy is supplied to the
planet, whereas the latter depends most strongly on surface
temperature. The runaway greenhouse occurs because water vapor is
a greenhouse gas and can therefore trap heat near the surface, and
hence is only a function of the absorptive properties of water and the
energy flux. The moist greenhouse occurs when the temperature at the
tropopause is large enough that water does not condense and is
therefore able to escape to the stratosphere where it can be
photolyzed. Increasing the planet's inventory of non-condensible
greenhouse gases has a small effect on the runaway greenhouse limit,
but can drive the moist greenhouse since the relative amount of
H$_2$O is lower. Thus, the runaway greenhouse is a more conservative
choice to demarcate the inner edge of the IHZ in the sense that it depends solely
on water, whereas knowledge of the atmospheric composition may not be
available to assess the likelihood of a moist greenhouse.

\begin{figure}
\includegraphics[width=6.5in]{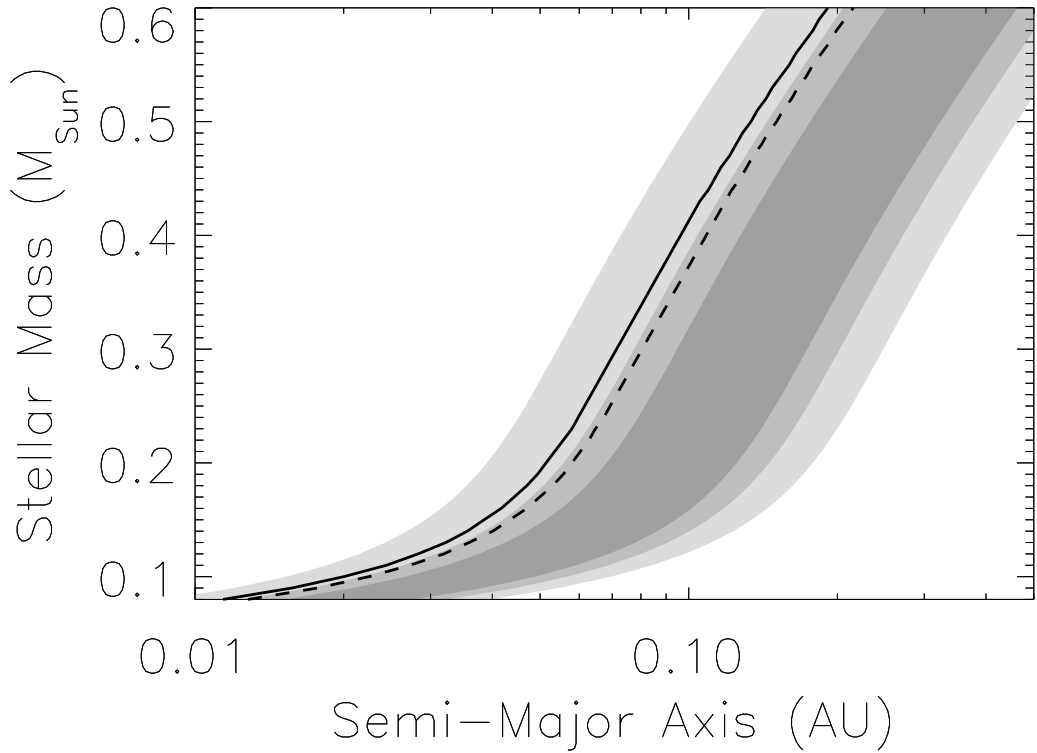}
\caption{\label{fig:hzcomp}Comparison of HZ boundaries for different
planetary masses and desiccating greenhouses. The shaded regions
represent the IHZ boundaries from \cite{Selsis07}: Dark gray assumes
no cloud coverage, medium gray 50\%, and light gray 100\%, and assuming 1 $\mearth$ planet. The
black curves represent the runaway greenhouse limit from
\cite{Pierrehumbert11}, with a planetary albedo of 0.49. The solid curve is for a 30 $\mearth$ planet, dashed for a 0.3 $\mearth$
planet. For these
calculations we used the stellar mass-radius relationship from \cite{BaylessOrosz06},
the mass-luminosity relationship from \cite{ReidHawley00}, and the terrestrial
mass-radius relationship from \cite{Sotin07}.}
\end{figure}

In Fig.~\ref{fig:hzcomp} we show the moist and runaway greenhouse for
wet planets in orbit around M dwarfs. See
Apps.~\ref{app:dg}, \ref{app:relations} and \ref{app:terrmr} for a discussion of IHZ
limits, stellar mass-radius and mass-luminosity relationships, and
terrestrial mass-radius relationships. In Fig.~\ref{fig:hzcomp}, the
grey regions are the limits of the moist greenhouse as presented in
\cite{Selsis07}. At the inner edge, \cite{Selsis07} find that
300~\wpmsq~triggers the moist greenhouse on a $1~\mearth$~planet, and,
from left to right, the limits assumed 100\%, 50\% and 0\% cloud cover
(for a $0.25~\msun$~ star these limits correspond to bond
albedos of 0.75, 0.49 and 0.23, respectively). The solid curve is the
runaway greenhouse limit \citep{Pierrehumbert10} for a
$30~\mearth$~planet and dashed for a $0.3~\mearth$ planet, both with
an albedo of 0.49 (compare to the medium grey). As expected (see App.~\ref{app:dg}) the
smaller planet's inner edge lies at larger semi-major axis than both
$1~\mearth$ moist greenhouse limit, which in turn lies farther out
than the $30~\mearth$ runaway greenhouse limit.

We have presented the classical description of the desiccation at the
inner edge of the habitable zone above. Various complications
are worthy of note. Water on a dry planet will get trapped at the
poles, making a moist or runaway greenhouse harder to achieve and
meaning that the inner edge of the HZ is nearer the star
\citep{Abe11}. We do not include that limit in Fig.~\ref{fig:hzcomp}
but we return to it in $\S$~\ref{sec:tv}.  Fig. \ref{fig:hzcomp}
assumes the planetary orbit is circular. However, for the many
exoplanets with (highly) eccentric orbits \citep{Butler06}, the total
irradiation over an orbit determines the annual-averaged surface
temperature
\citep{WilliamsPollard02} and pushes the
IHZ boundaries out by a factor of $(1 - e^2)^{- 1/4}$
\citep{Barnes08_hab}.

%
%
%
%

\section{Tidal Heating \label{sec:tides}}

Tidal heating is responsible for the volcanic activity on Io
\citep{Peale79}, and probably the geysers of Enceladus \citep{Hansen06,Porco06,Hurford07}. Tidal theory
has a long and established body of work, however tidal processes
remain poorly understood. The difficulty lies in the complexity of the
energy dissipation processes and the very long timescales associated
with tidal evolution. The Earth-Moon system is the most accurately
studied, with lunar laser ranging providing a precise measurement of
the recession of the Moon due to tides of $\sim 38$ mm/yr
\citep{Dickey94}, as well as direct measurements of the locations of dissipation in the ocean \citep{EgbertRay00}. However,
using the currently estimated tidal dissipation parameters to
extrapolate backwards in time predicts
that the Moon was at the Earth's surface about $\sim 2$ Gyr ago
\citep{MacDonald64}, which contradicts the standard impact model for the lunar origin (note that numerous issues remain such as the
origin of the Earth and Moon's obliquities \citep{ToumaWisdom94} and
the perturbations from other planets \citep{Cuk07}). The deceleration
of Io's orbital velocity has also been tentatively detected and seems
broadly consistent with tidal theory
\citep{AksnesFranklin01,Lainey09}. However, these data lie in the same
regime, that of nearly-circular orbits. Exoplanets have been found
with extremely eccentric orbits, so we can only extrapolate from our
Solar System cautiously. The details of tidal theory are complicated,
and hence we relegate the discussion to App.~\ref{app:tides}.

Two end-member models of tides theory have been applied to exoplanets
and the bodies of our Solar System: One assumes the energy dissipation
gives a constant phase lag in the periodic distortion (CPL), and the
other assumes a constant time lag (CTL)
\citep{Greenberg09}. While simple and linear, such models are probably commensurate with the dearth of information we
have about exoplanet interior processes. More complicated models have
been constructed, and they reproduce the above models for certain
choices of internal composition, structure, and energy transport
\citep[\eg][]{Henning09}.  Thus, the CPL and CTL models can
provide important and accurate insight into the tidal evolution of exoplanets.

These models converge at $e = 0$, and have been shown to be nearly
identical for $e
\lsim 0.2$ \citep{Leconte10}, and when using Eq.~(\ref{eq:qtau}). However, for $e \gsim 0.3$, they diverge
significantly. We urge caution when interpreting results in the upcoming
sections which allow $e$ to be as large as 0.8. We include this range primarily
for illustrative purposes, and as a baseline for any future work which
may include non-linear effects.

The tidal heating of a body is provided in Eqs.~(\ref{eq:E_tide_cpl})
and (\ref{eq:E_tide_ctl}) for the CPL and CTL models,
respectively. Averaging the heating rate of the entire planetary
surface gives the surface energy flux due to tides, $F_{tide}$. In
those equations, the strength of the tidal effects is parametrized as a ``tidal quality factor'' $Q$ (CPL) or ``tidal time lag'' $\tau$
(CTL), which are notoriously difficult to measure or estimate from first principles. The Earth's current
values are $Q = 12$~\citep{Yoder95} and $\tau = 638$~s
\citep{Lambeck77,NeronDeSurgyLaskar97}, respectively. However, as
noted above, these values predict too short a lifetime of the
Moon. This discrepancy has led to the notion that the Earth's response
to lunar tides has evolved with time, possibly due to changing size,
shape and seafloor topography of the oceans, affecting the ocean currents
response to the tidal potential. Measurements of the dry bodies in our
Solar System have found that their $Q$s tend to cluster around 100,
see App.~\ref{app:tides}.

Recent satellite observations of the Earth have revealed the locations
of tidal dissipation in our oceans \citep{EgbertRay00}. Tides force water
through shallow seas and straits causing energy dissipation. In the
open ocean, tidal dissipation is probably a non-linear process in
which currents are disrupted by seafloor topography. The presence of
the ocean provides more opportunity for tidal dissipation than on dry
planets, and we assume that most dissipation on habitable
exoplanets will also occur in their oceans. Therefore, for the following calculations we assume modern
Earth-like planets with $Q = 10-100$ or $\tau = 64-640$~s.

%
%
%
%

\section{Tidal Venuses \label{sec:tv}}

\begin{figure}
\includegraphics[width=5in]{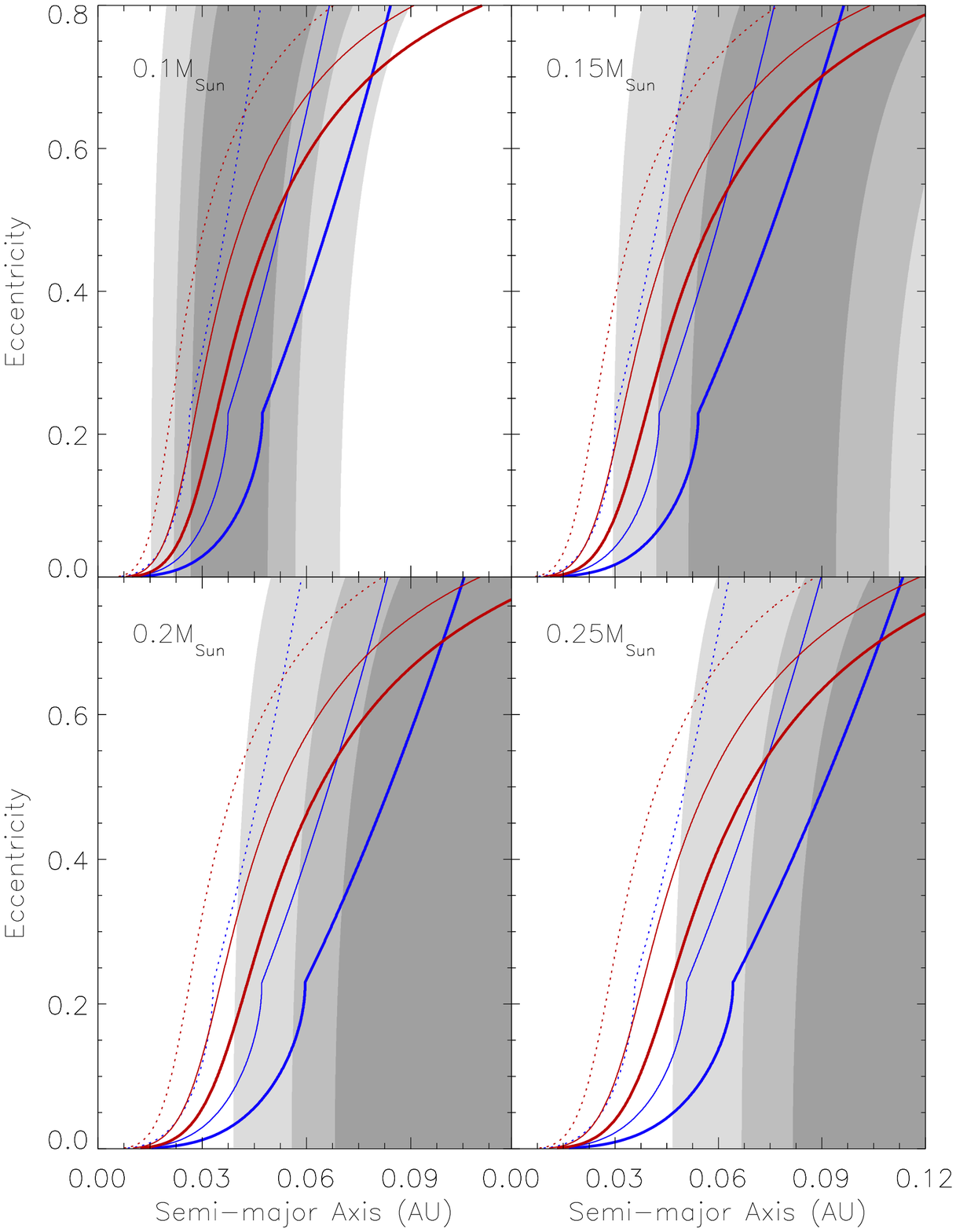}
\caption{\label{fig:tv4}Parameter space of the Tidal Venus. The top left panel is a $0.1~\msun$ star, top right $0.15~\msun$, bottom left $0.2~\msun$, and bottom right
$0.25~\msun$. The grayscale represents the \cite{Selsis07} IHZ
boundaries: From lightest to darkest gray, the cloud coverage is
100\%, 50\% and 0\%, respectively.  The colored curves mark
where $F_{tide} = F_{crit}$. Red curves assume the CTL model,
blue the CPL. Solid curves assume the \citep{Pierrehumbert11} runaway
greenhouse model, and dotted the dry world model of
\citep{Abe11}. Thick lines are for a $10~\mearth$ planet, and thin for
$1~\mearth$. Tidal Venuses lie to the left of these curves.}
\end{figure}

We computed tidal heating rates for a range of planetary, stellar
and orbital parameters and find that tidal heating can be strong
enough on some planets to trigger a runaway greenhouse. In
Fig.~\ref{fig:tv4} we show the configurations that predict this state
around 4 different hypothetical M dwarfs. The IHZ boundaries are the
moist greenhouse limits from \cite{Selsis07} and with the same format as Fig.~\ref{fig:hzcomp}. The colored curves mark
where $F_{tide} = F_{crit}$. Red curves assume the
CTL model, blue the CPL with discrete rotation states, see App.~\ref{app:tides}.1. Solid curves assume the
\cite{Pierrehumbert11} runaway  greenhouse model, and dotted the dry world model of \cite{Abe11}, \cf
Fig.~\ref{fig:rg} in App.~\ref{app:dg}. For these latter worlds,
we choose $Q = 100$ as they do not have oceans. Thick lines are for a
$10~\mearth$ planet, and thin for $1~\mearth$ (\cite{Abe11} only
considered a $1~\mearth$ planet). Note that our choice for the
relationship between $Q$ and $\tau$ influences which model predicts
more heating. If we had chosen the same relationship as in
\cite{Heller11}, we would have found the CTL model predicts
more heating than CPL, \cf their Fig.~5.

For the lowest mass M dwarfs, the HZ is significantly
reduced due to tidal heating. For masses larger than $0.25~\msun$, a
tidal greenhouse in the IHZ is only possible on large mass planets
with very large eccentricities. Note that our model predicts a tidal
greenhouse at low eccentricities where tidal theory is most likely
valid. Figure \ref{fig:tv4} shows that a planet may have a climate
catastrophe due to tide-driven overheating even if it is far enough from the star
that stellar radiative heating alone would not preclude habitability.

The difference between curves representing equal mass planets is due to the
frequency dependence of the CTL model. For the Earth, the frequency is
the mean motion of the lunar orbit, but a planet at $a = 0.035$ AU, \ie
the middle of the HZ, orbits in about 1 week. One could formally
adjust $\tau$
so that it is equivalent to a $Q$ of 10 (\cite[see
\eg][]{Matsumura10,Heller11} and Eq.~[\ref{eq:qtau}]) and then the
curves would lie in a similar location. The unknown tidal
response of terrestrial bodies and the absence of an unambiguous
translation from the CPL to the CTL model motivates our choice of
adopting present-Earth values for all planets at all frequencies.

\begin{figure}
\includegraphics[width=5.5in]{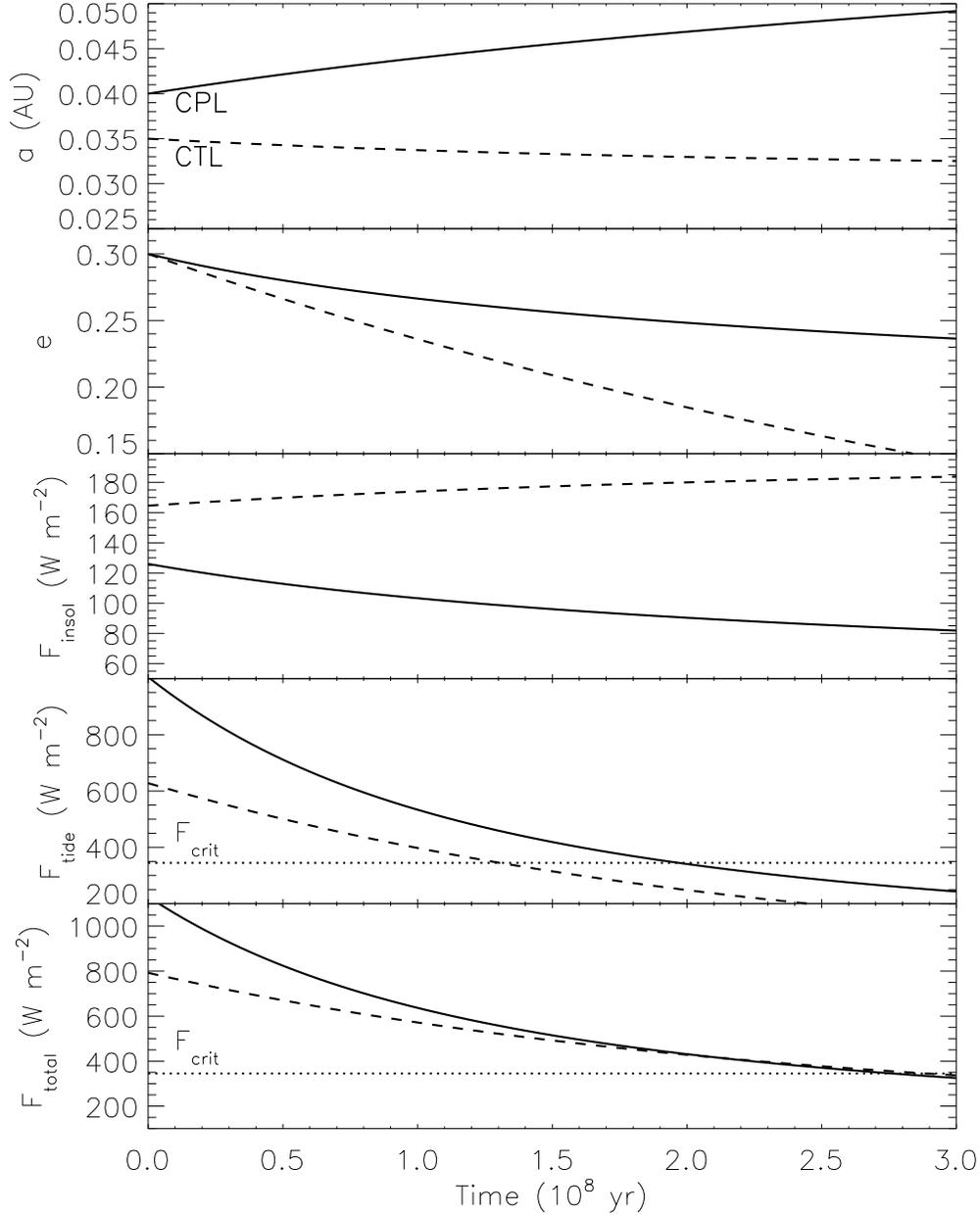}
\caption{\label{fig:evol} Evolution of a $10 \mearth$ planet orbiting  a 
$0.1 \msun$ star with an initial orbit of $a = 0.04$~AU, $e = 0.3$. {\it  
Top:} Semi-major axis evolution. {\it Top middle:} Eccentricity  
evolution. {\it Middle:} Insolation evolution. {\it Bottom Middle:} Tidal 
heat flux evolution. {\it Bottom:} Total surface heat flux (insolation + 
tidal).}
\end{figure}

Regions to the left of the colored curves in Fig.~\ref{fig:tv4} can
produce Tidal Venuses if the planets remain there longer than
$t_{des}$. In Fig.~\ref{fig:evol}, we show the evolution of two
example systems consisting of a $10 \mearth$ planet orbiting a $0.1 \msun$ star,
using both the CPL and CTL models. For the CPL case, the planet begins
with semi-major axis $a = 0.04$ AU, and $e = 0.3$. The planet is assumed to be
spin-locked and with zero obliquity. This orbit is
toward the outer edge of the IHZ and with the typical eccentricity of
known exoplanets. For the CTL case, everything is the same,
except the initial $a$ is 0.035~AU, as the Tidal Venus region lies
closer to the star. The top panel in Fig.~\ref{fig:evol} shows the evolution
of $a$, the next panels down show $e$, then insolation $F_{insol}$,
then tidal heat flux $F_{tide}$,  and finally the sum of the insolation
and tidal heat flux, $F_{total}$.  The tidal heat flux generated in
the planet in the CPL case has decreased to $F_{crit} = 345$~\wpmsq~at
200 Myr, and the sum of tidal heating and insolation reaches
$F_{crit}$ at 275 Myr. In the CTL case, $F_{tide} = F_{crit}$ at
130~Myr, and $F_{total} = F_{crit}$ at 275~Myr.  We conclude that planets
orbiting low-mass stars on eccentric orbits may experience tidal
greenhouse conditions long enough to become uninhabitable.

The CPL model predicts $a$ will increase through angular
momentum transfer with the star, for which we assume a rotation period
of 30 days, a tidal $Q$ of $10^6$, a radius determined by the
\cite{ReidHawley00} relation, and both the radius of gyration
and Love number of degree 2 are 0.5. However, this orbital expansion
is a result of the discrete nature of the CPL model, which only
includes 4 ``tidal waves,'' see App.~\ref{app:tides}.1. In
Fig.~\ref{fig:evol} the planet initially spins 3 times per 2 orbits,
and hence $\varepsilon_{1,1} = 0$, see Eq.~(\ref{eq:epsilon}),
eliminating one of the terms in d$a$/d$t$. Note that once the
eccentricity drops below $\sqrt{1/19}$, the spin rate becomes
synchronous (see App.~\ref{app:tides}.3), and then $\varepsilon_{0,1}
= 0$ but $\varepsilon_{1,1} = 1$, and then d$a$/d$t < 0$. For this
example, that transition occurs at 375~Myr, and is not shown in
Fig.~\ref{fig:evol}. This point illustrates the complexity inherent in
this commonly-used tidal model.

%
%
%
%

\section{Constraining Observed Planets \label{sec:history}}

The previous section demonstrated that planets may become desiccated
by tidal heating, but in many cases we will be more interested in the
possibility that such a condition developed on a planet that we have
discovered. As tides tend to circularize orbits, we may find a planet
in the IHZ with low eccentricity that experienced a tidal greenhouse
early on and is hence currently uninhabitable. In this section we
describe how to evaluate a known planet's probability for habitability
based on past tidal heating.

In order to model the tidal heating history of an exoplanet, we
require knowledge of $M_p$, $R_p$, $a$, $e$,
$M_*$, $R_*$ and age. Exoplanets are predominantly discovered by RV
and transit studies that can provide $a$ and $e$, while the others can
be modeled from the stellar spectrum. If some of these values are
unknown, we can often use scaling relations to estimate them, see
App.~\ref{app:relations}--\ref{app:terrmr}. From this information, one
can estimate the tidal heating history of a planet, using the tidal
evolution equations (App.~\ref{app:tides}), and hence estimate the
probability that the planet has lost its water.

As an example, consider the hypothetical situation in which a
terrestrial-scale planet has been discovered in orbit about an M
dwarf. Further, the planet lies near the inner edge of the IHZ, and
has a low eccentricity. To evaluate past tidal heating, we created 4
systems with different stellar and planetary properties, applied
reasonable uncertainties to each observable properties, chose
non-observable properties via appropriate scaling laws, and modeled
the system's history. In particular, we assume that the planet's mass
lies between 1 and 5 $\mearth$, has an eccentricity less than 0.1, a
tidal $Q_p$ ($\tau_p$) in the range 10 -- 100 (640 -- 64 s), and is
tidally-locked. We assume the star's age lies between 2 and 8 Gyr, and
$Q_*$ ($\tau_*$) in the range $10^5$ -- $10^{10}$ (1 -- $10^{-4}$ s)
and $t_{des} = 10^8$ years. We used the same scaling relations as
in $\S$~\ref{sec:tv}. For the $0.1~\msun$ case, $a$ was chosen in the range
[0.029,0.031] AU, for $0.15~\msun$ it was [0.055,0.065] AU, for
$0.2~\msun$ it was [0.085,0.095] AU, and for $0.25~\msun$ it was
[0.095,0.105] AU. For each parameter the uncertainty distribution was
uniform in the quoted range, except for $e$ which was chosen uniformly
in the range $-5
\le \textrm{log}_{10}(e) \le -1$. We then randomly determined the system
parameters in these ranges and integrated the tidal history backward
in time, using the models presented in App.~\ref{app:tides}, for the
randomly chosen age of the system. We ignore the possibility of
spin-orbit resonance capture, which would dramatically alter the
history. For each stellar mass and tidal
model, we simulated 30,000 possible configurations.

\begin{figure}
\includegraphics[width=4.3in]{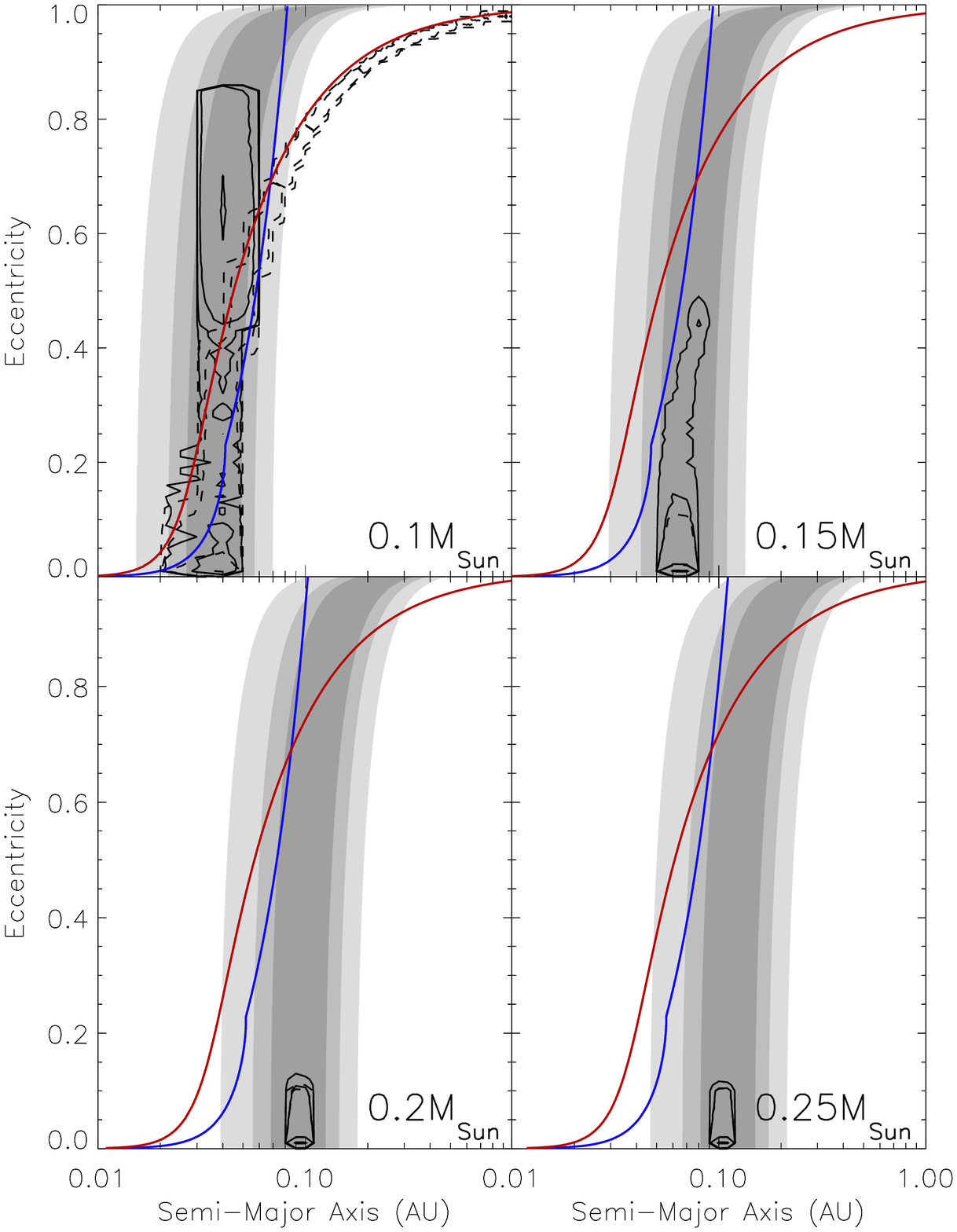}
\caption{\label{fig:history}Orbits of hypothetical planets  around M dwarfs after $t_{des}$. The top left panel is a $0.1~\msun$ star, top right $0.15~\msun$, bottom left $0.2~\msun$, and bottom right
$0.25~\msun$. The grayscale represents the \citep{Selsis07} IHZ
boundaries with the same format as Fig.~\ref{fig:tv4}. For reference, the red and blue curves show the tidal greenhouse limit for a $2.5~\mearth$ planet with
$Q_p = 10$ (CPL model, blue curve) or $\tau_p = 640$ s (CTL model, red
curve). Contours denote levels of constant probability density (for
densities 50\%, 90\% and 99\% of the peak value) for the
initial orbit of the planet: solid corresponds to the CPL model
(compare to blue curve), dashed to CTL (compare to red curve). In the
bottom two panels there has been negligible orbital evolution for
either tidal model.}
\end{figure}

In Fig.~\ref{fig:history} we show our results graphically. As we are
considering a range of masses, we choose to represent the runaway
greenhouse flux with a $2.5~\mearth$ planet, \ie the solid curves show
where $F_{tide} = F_{crit} = 309$~\wpmsq. The three contours show the
probability density for the planet's location $t_{des} = 10^8$ years
{\it after the system's formation}. The contours show where the probability
has dropped by 50\%, 90\% and 99\% of the peak value, solid contours
are for the CPL model, dashed for the CTL. If the probability contours
intersect, or even come close to, the colored $F_{tide} = F_{crit}$
curves, then the planet may be a ``Habitable Zone Venus,'' a planet
that appears habitable by the IHZ metric, but is probably more
Venus-like than Earth-like. 

The shapes of the regions in Fig.~\ref{fig:history} are due to
our parameter choices and the assumptions implicit in the two tidal
models. The CPL model does not include eccentricity terms to as high
an order as the CTL model, and hence the CTL model predicts more
evolution at large $e$. In the top left panel, this effect is seen
clearly, as the most likely orbits at 100~Myr in the CPL model are at
lower semi-major axis than the CTL model. We stress that this example
is purely hypothetical, and the actual shapes of the contours could be
very different for actual systems.

For the $0.1~\msun$ case (top left), the planet may have spent enough
time in the tidal greenhouse to be uninhabitable. For larger mass
stars, however, the danger of tidal desiccation is
smaller. In the $0.15~\msun$ case, the peak in the CPL probability
density at $a = 0.08$, $e = 0.45$ represents 1\% of our
simulations. Although the contours do not cross the CPL runaway
greenhouse (blue) curve, they do come close, and hence there is a
small chance that our putative candidate is a super-Venus, especially
if we allow for absorption of stellar radiation.

However, for $M_* > 0.15 \msun$, planets with low eccentricity
probably have always had low eccentricity, \ie the evolution was
negligible. This sharp contrast between 0.1 and 0.2 $\msun$ occurs
because of the steep dependence of tidal heating on $a$. At larger
stellar masses, the circular IHZ has been pushed away from the reach
of fatal tidal heating.

This experiment is purely illustrative with arbitrary parameters and
uncertainty distributions. If a terrestrial planet was discovered in
the IHZ of a very low mass star, then this methodology could be performed in
order to characterize its potential to support life. One could derive
a value of $t_{des}$ tailored to the primary, and determine the
probability that the planet spent more than that in a tidal greenhouse
state and hence is uninhabitable. In principle, one should also allow
a range of albedos and include radiation to determine the amount of
time that the total surface energy flux, $F_{tot} = F_{tide} +
F_{insol}$, is larger than $F_{crit}$, but we leave such an analysis
for future work.

%
%
%
%

\section{Application to Gl 667C\label{sec:gl667c}}

Two or more planets orbit the $0.3~\msun$ star Gl 667C
\citep{Bonfils11,AngladaEscude12,Delfosse12}. Planets c and d
appear to lie in the IHZ, while a third planet, b, lies
interior. Planet c is at least $4.5~\mearth$ and orbits near the
inner edge of the IHZ. Planet d, which is weakly detected in both
studies, lies near the 50\% cloud cover outer boundary. At 0.23 AU,
planet d is too far from the star to be subjected to strong tidal heating, at least
if it is tidally locked. \cite{AngladaEscude12} and \cite{Delfosse12}
propose different solutions to the system with the former setting c's
eccentricity to 0, but stating that it is only constrained to be
$<0.27$, while the latter assign its eccentricity to be $0.34 \pm
0.1$. The minimum mass estimates are almost identical at
$4.25~\mearth$~\citep{AngladaEscude12} and
$4.5~\mearth$~\citep{Delfosse12}. Here, we use the \cite{AngladaEscude12}
solution, but note that using data from \citep{Delfosse12} does
not change our results.

\begin{figure}
\includegraphics[width=5in]{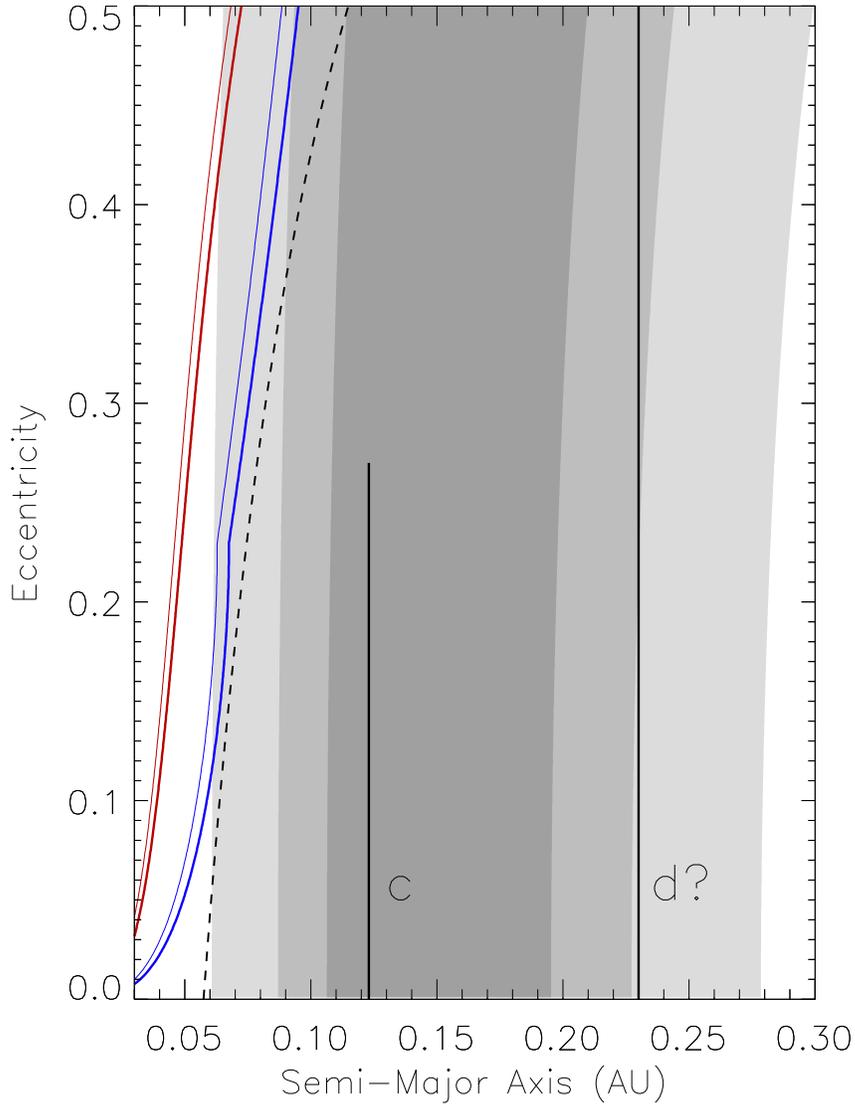}
\caption{\label{fig:gl667c}Comparison of the IHZ to the tidal
greenhouse limits for Gl 667C c, in a similar format as
Fig.~\ref{fig:tv4}. Here the thin lines correspond to a $4.5~\mearth$
planet, and thick to $9~\mearth$. The vertical black lines correspond
to the uncertainty in eccentricity of c and d (the latter's existence remains
uncertain). The dashed, black curve represents where c's and b's orbits
cross, and hence the region to the left is dynamically unstable.}
\end{figure}

In Fig.~\ref{fig:gl667c}, we show the system in the same format as
Fig.~\ref{fig:tv4}. As this planet was detected via radial velocity
data, its true mass is unknown. We consider two possibilities here, an
edge-on geometry in which its actual mass is $4.5~\mearth$ and an
inclined case in which its actual mass is doubled, $9~\mearth$. In
Fig.~\ref{fig:gl667c}, the thin lines correspond to the minimum mass,
thick to twice-minimum. The vertical extent of the line
corresponds to the uncertainty in eccentricity. The orbit of c is marked by the vertical black
line at $a = 0.123$~AU, and d at 0.23~AU.

The CTL model (red curves) barely intersect the 100\% cloud cover IHZ
at large $e$. The CPL model penetrates the IHZ more significantly,
but still does not reach the orbit of c. Thus, planet c is not
currently experiencing a tidal greenhouse, assuming no obliquity and
pseudo-synchronous rotation. We also see that d is completely safe
from the tidal greenhouse.

But could planet c have been in the tidal greenhouse and be
desiccated today? The answer is almost assuredly no. In this case, we
can appeal to the orbital architecture, rather than running a suite
of Monte Carlo simulations as in $\S$~\ref{sec:history}. The dashed,
black line of Fig.~\ref{fig:gl667c} showed where the orbit of c
crosses the orbit of b. Such a configuration is as close as the two
planets could possibly be and remain stable, assuming no mean motion
resonances, which are not detected in this system
\citep{AngladaEscude12}. This black curve is always exterior to the
tidal greenhouse curves out to the current orbit of c. Therefore, we
may safely conclude that c was never in a tidal greenhouse due to its
orbit. Furthermore, the planet should have become tidally-locked
within $10^7$ years \citep{Heller11}, which is much less than
$t_{des}$. Hence any initial burst of tidal heating due to
non-equilibrium rotation or obliquity was too short to sterilize this
planet. Gl 667C c remains a habitable planet candidate.

\section{Discussion\label{sec:disc}}

The previous results show where tidal heating can limit habitability
for planets orbiting low mass stars. However, many phenomena
complicate the process and could alter our findings. Here we discuss
the results of the previous sections in light of our simplifying
assumptions, observational requirements, and theoretical limitations.

We find that Gl 667C c probably did not lose its water to tidal
heating because interactions with other planets prevent its
eccentricity from being large enough to trigger the tidal
greenhouse. As more potentially habitable planets are discovered
around low mass stars, a similar analysis as in $\S$~\ref{sec:history}
should be undertaken in order to assess the possibility that the
planet could in fact be dehydrated. As we may only be able to
spectroscopically characterize a few planets with the \textit{James
Webb Space Telescope} \citep{Seager09,KalteneggerTraub09},
prioritization of targets is crucial, and past and present tidal
heating will help determine the best planet to observe.

The timescale for $e$ to decay may be smaller than the timescale for
in situ terrestrial planet formation
\citep{LecarAarseth86,WetherillStewart89,Lissauer93,KokuboIda98,Raymond07}.
On the other hand, terrestrial planets could be pushed into such a
position by a migrating gas giant in about that timescale
\citep{Raymond06,Mandell07}. It is therefore natural to wonder if such
planets can even form, as the tidal effects could suppress planet
formation, or damp out the eccentricity of a protoplanet before it is
massive enough to support clement conditions. Several scenarios
suggest that fully-formed planets can be Tidal Venuses. Orbital
instabilities can excite $e$. Planet-planet scattering and divergent
resonance crossing appear to play a role in sculpting many planetary
systems, including our own
\citep[\eg][]{WeidenschillingMarzari96,Tsiganis05,Nesvorny11}. These phenomena
can develop long after planet formation has occurred. Recently it has
been suggested that exoplanetary systems in resonance appear
systematically younger than the general population, suggesting that
instabilities can even occur after many billions of years
\citep{KoriskiZucker11}. Therefore, we conclude that tidal greenhouses
are plausible.

The role of oceans in the tidal dissipation process is clearly very
important, yet also very poorly understood. We have used up-to-date
information regarding the dissipation of energy in the Earth's ocean
and assumed that terrestrial exoplanets will behave similarly, yet
many issues remain outstanding. We have ignored the role of inertial
waves, which provide additional heating of aqueous mantles of icy
satellites \citep{Tyler08,Tyler11}. Therefore the tidal heating values
we obtained may, in fact, be too low, further increasing the threat to
habitability. Regardless, our ranges for $Q$ or $\tau$ spanned an order of magnitude, but a wider range of values is still possible, as
they are complex functions of ocean depth, ocean floor topography and
the shape of any continents or islands that may be present. As
``exobathymetry'' seems a distant dream, tidal dissipation in
Earth-like worlds will remain mysterious for the foreseeable future.

Further complicating the situation, oceans may evaporate long before
$t_{des}$, potentially leading
to a cyclical process of evaporation and precipitation: After the oceans
disappear, $Q$ increases and the tidal heat decreases, causing the
planet to drop out of the tidal greenhouse, so
the water rains out, reforming the oceans and lowering $Q$
again. Whether such a cycle exists is pure speculation, but we note that
an analogous situation can occur with the classic IHZ, where evaporated
water forms clouds that increase the albedo, which in turn lowers the
upward long-wavelength flux from the surface, and the planet then
drops out of the runaway greenhouse. Our choices for
$Q$, $\tau$, $F_{crit}$ and $t_{des}$ can be revised as new
observations provide firmer constraints. Of these, $t_{des}$ needs the
most work, as it is a function of poorly constrained host properties,
and challenging planetary escape processes, see \eg \cite{Tian09}

A Tidal Venus is an extreme case of tidal heating, over two orders of
magnitude more powerful than on Io. We have therefore made a considerable
extrapolation. Perhaps oceans and/or mantles adjust to the increased
heating and fail to reach that flux. Our choice for $Q$ and $\tau$
imply most dissipation occurs in the ocean. For the solid interior,
\cite{Behounkova11} find that tidal heating at about $F_{crit}$ leads
to a ``thermal runaway'' for planets that transport internal energy
through convection. Although they did not consider the possibility of
advection via volcanism, the geophysics of Tidal Venuses require
closer scrutiny. We are unaware of any research exploring the physical
oceanography of planets undergoing strong tidal heating, suggesting it
is an interesting topic for future research.

Our treatment ignored mutual gravitational interactions between
planets or large satellites, which can significantly alter the
evolution. Gravitational perturbations can pump up eccentricities and
obliquities to non-zero values and may be able to modify the spin
period. From Fig.~\ref{fig:tv4}, we see that planets near the inner
edge of the HZ may be in a tidal greenhouse for eccentricities less
than 0.02. Other bodies in the system can easily perturb
eccentricities to larger values, hence planets in multiple systems may
be especially susceptible to a tidal greenhouse, as this driven
eccentricity can be maintained for arbitrarily long timescales. As
individual systems are discovered, this point should be revisited.

On the other hand, orbital stability arguments could allow us to
preclude the tidal greenhouse in multiplanet systems, as for Gl
667C. When the eccentricity required for the tidal greenhouse is so
large that the system is unstable, then one can safely exclude the
Tidal Venus state. Hence, the presence of additional companions
provides critical information when assessing habitability.

Many other critical phenomena were also left out such as atmospheric
erosion by flaring, stellar activity and magnetic dynamo generation
\citep{Lammer07,Khodachenko07,Tian09,Segura10,Lammer10,DriscollOlson11}. Tidal heating
provides an interesting counterbalance to atmospheric erosion as it
may increase the outgassing rates and maintain a permanent
atmosphere. The outgassing and escape need to remain in a balance as
wild pressure and density fluctuations will undoubtedly alter the
biosphere, but in principle, tidally-driven outgassing could reduce
the danger of atmospheric removal. On the other hand, more intense
outgassing without commensurate draw down by processes like the
carbonate-silicate cycle could increase the threat of a moist
greenhouse. Magnetic fields may slow down atmospheric loss, but tidal
heating may decrease the dynamo. If magnetic fields are generated by
convection in the outer core between a hot inner core and a cool
mantle \citep{OlsonChristensen06,DriscollOlson11}, then tidal heating
of the mantle may suppress magnetic field generation
\citep{Stevenson10}. This issue has not yet been explored for
tidally-heated exoplanets, but it could play a major role in
habitability. Future work should couple outgassing and escape rates in
order to determine how the two interact.

\section{Conclusions \label{sec:concl}}

We have shown that tidal heating of some exoplanets may exceed the
threshold of the runaway greenhouse, the traditional inner edge of the
IHZ. We find that for stars with masses $\lsim 0.3~\msun$, planets in
their IHZs with low eccentricity, can be uninhabitable regardless of
insolation. We have thus fundamentally revised the HZ boundaries for
planets on eccentric orbits. Unlike insolation from main sequence
stars, tidal heating at a desiccating greenhouse level may drop off
rapidly, but not so rapidly as to preclude the possibility that a
planet's entire inventory of water can be lost permanently, see
Fig.~\ref{fig:evol}. These planets will be uninhabitable regardless of
future tidal heating, \ie a planet found with minimal tidal heating
today may still have experienced sufficient heating for sufficient
duration to render it uninhabitable. Additional planetary
companions are important: They can drive eccentricity and sustain a
tidal greenhouse, or they can be used with stability arguments to rule
out an early tidal greenhouse.

Traditionally, habitability models have focused on insolation,
implying the star is the most important aspect of habitability. We
have shown that in some circumstances tidal effects are more important
in determining the inner edge of the HZ. Planetary habitability is a
function of the star, the planet, and the planetary system
\citep{Meadows12}. The \cite{Kasting93} IHZ has served
well as a guide, but is insufficient. Combining all the
processes relevant to habitability into a single model is a daunting
challenge to say the least, but a proper assessment of a planet's
potential for habitability relies on a wide diversity of properties,
some of which will not be observable any time soon. Nevertheless, the
prospect of identifying an inhabited planet is strong
motivation. Moreover, the high cost in time, money and resources
required to establish a planet as inhabited demand that we use these
resources efficiently. In this study we have compiled numerous tidal
processes and empirical relationships so that at least the tidal
effects predicted by linear theory may be applied to terrestrial
planets in the IHZs of low luminosity hosts.

\medskip

This work was funded by NASA Astrobiology Institute's Virtual Planetary
Laboratory lead team, under cooperative agreement No. NNH05ZDA001C. RB
acknowledges additional funding from NSF grant AST-1108882. We are
also grateful for stimulating discussions with Richard Greenberg, Norm 
Sleep, Sean Raymond, and Andrew Rushby.

\bibliography{TidalVenus_AsBio}

%
%
%
%

\appendix

%
%
%
%

\section{Relations for the inner edge of the habitable zone \label{app:dg}}


The critical flux for the runaway greenhouse depends weakly on the
acceleration due to gravity, $g$, in the planetary atmosphere. Flux is
emitted to space from the level of optical depth unity. The optical
depth of a layer of absorbing gas depends to first approximation on
the mass of the layer, so with higher $g$ the pressure at the base of
a constant mass layer will be higher. For the Simpson-Nakajima limit,
the $T$-$p$ structure of the atmosphere is the saturation vapor
pressure curve, so higher $p$ implies higher $T$ and permits a higher
critical flux. This positive relationship is offset to some extent by
pressure-broadening of absorption, so at higher $g$ (and higher $p$),
less mass of water is required to provide an optical depth of unity
\citep{Pierrehumbert10,GoldblattWatson12} .

\begin{figure}
\includegraphics[width=6.5in]{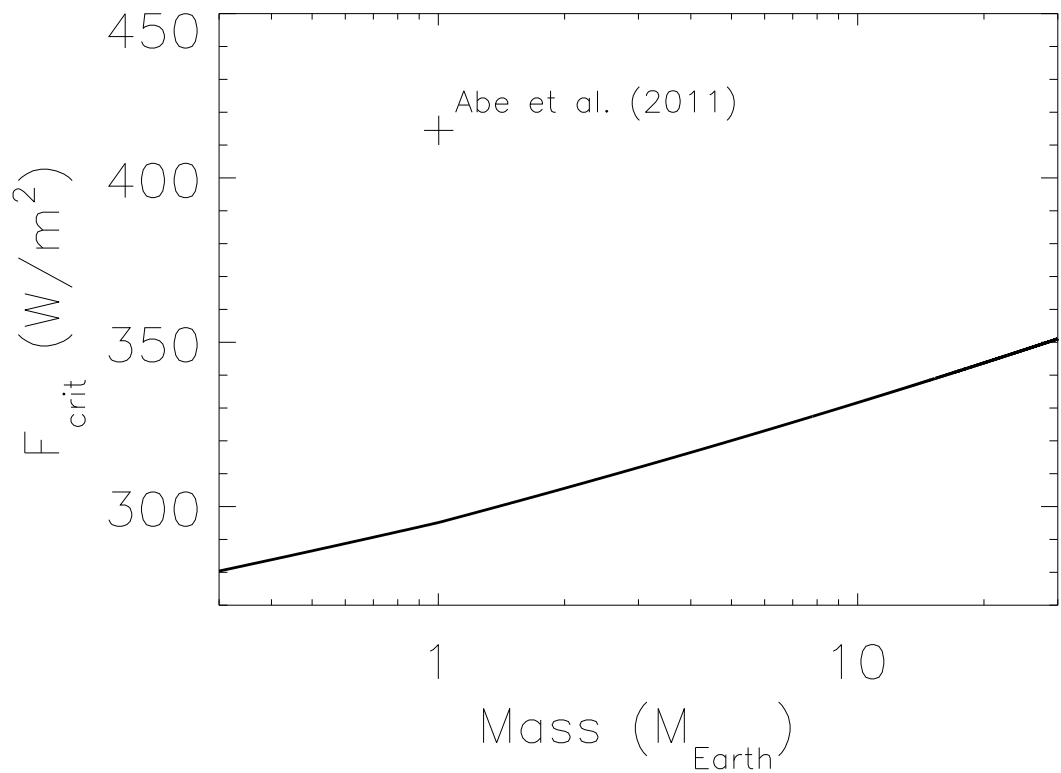}
\caption{\label{fig:rg}Surface flux required to trigger a runaway greenhouse ($F_{crit}$)
as a function of planet mass. The solid line is the relation for wet
planets from
\cite{Pierrehumbert11}, and the
cross is the limit from \cite{Abe11} for a dry planet.}
\end{figure}

\citet{Pierrehumbert10} derives a semi-analytical relationship for the critical 
flux due to the Simpson-Nakajima limit:
\begin{equation}
\label{eq:RG_P10}
F_\mathrm{crit} = A \sigma \left( \frac{l}{ R  \ln\left(P_\star
\sqrt{\frac{\displaystyle \kappa}{\displaystyle 2P_0 g}}\right)     } \right)^4
\end{equation}
where $l$ is the latent heat capacity of water, $R$ is the universal
gas constant, $P_0$ is the pressure at which the line
strengths are evaluated, $g$ is the gravitational acceleration in the planetary atmosphere. $\kappa$ is a gray absorption coefficient and
$A$ is a constant of order unity; \citet{Pierrehumbert10} fits these to numerical runs of a radiative transfer code with $\kappa = 0.055$ and $A = 0.7344$. 
\begin{equation}
P_* = P_\text{ref}e^\frac{l}{RT_\text{ref}},
\label{eq:pstar}\end{equation}
where $T_{ref} = 273.13$\,K and $P_{ref} = 610.616$\,Pa are points on the saturation vapor pressure curve of water, taken as the triple point. This calculated limit is shown in Fig. \ref{fig:rg}.

For convenience, we use an empirical relationship for the moist
greenhouse limit derived by \cite{Selsis07}:
\begin{equation}
\label{eq:lin}
l_{in} = (l_{in\odot} - a_{in}T_* - b_{in}T_*^2)\Big(\frac{L_*}{L_\odot}\Big)^{1/2
}.
\end{equation}
The outer edge (not due to a desiccating greenhouse) can be expressed
in a similar format as
\begin{equation}
\label{eq:lout}
l_{out} = (l_{out\odot} - a_{out}T_* - b_{out}T_*^2)\Big(\frac{L_*}{L_\odot}\Big)^{1/
2
}
\end{equation}
In these equations $l_{in}$ and $l_{out}$ are the inner and outer edges of the IHZ,
respectively, in AU, $l_{in\odot}$ and $l_{out\odot}$ are the inner
and outer edges of the IHZ in the solar system, respectively, in AU,
$a_{in} = 2.7619 \times 10^{-5}$ AU/K, $b_{in} = 3.8095 \times
10^{-9}$ AU/K$^2$, $a_{out} = 1.3786 \times 10^{-4}$ AU/K, and
$b_{out} = 1.4286 \times 10^{-9}$ AU/K$^2$ are empirically determined
constants,  and $L_*$ and $L_\odot$ are the stellar and solar
luminosity, respectively.
$T_* = T_{eff} - 5700$\,K, where $T_{eff}$ is the ``effective
temperature'' of the star
\begin{equation}
\label{eq:Teff}
T_{eff} = \big(\frac{L_*}{4\pi\sigma R_*^2}\big)^{\frac{1}{4}},
\end{equation}
where $R_*$ is the stellar radius and $\sigma$ is the
Stefan-Boltzmann constant. Usually, however, the IHZ is couched
in terms of the stellar mass $M_*$ and $a$, and hence
relationships between $L_*$, $M_*$ and $R_*$ are needed to produce
such an expression. Empirical and theoretical relations
between these properties are reviewed in App.~\ref{app:relations}.

The values of $l_{in\odot}$ and $l_{out\odot}$ are therefore the key
parameters in the identification of the edges of the IHZ. We consider
three criteria identified in \cite{Selsis07}: 1) 0\% cloud cover, 2)
50\% cloud cover, and 3) 100\% cloud cover. \cite{Selsis07} give
values of $l_{in\odot}$ of 0.89, 0.72, and $\sim 0.49$ AU for the
three possibilities, respectively. For the outer edge, \cite{Selsis07}
give values of $l_{out\odot}$ of 1.67, 1.95, and 2.4 AU, for the three
cloud cover models, respectively. We arbitrarily choose the 50\% cloud
cover case to be the limits of the IHZ. This choice is the middle of
the possibilities, and roughly corresponds to the cloud cover on the
Earth.

However, as \cite{Selsis07} note, ``the effect of the spectral type on
the albedo, included in [these equations] as a quadratic function of
($T_\text{eff} -5700$), was estimated only for a cloud-free
atmosphere. Since the reflectivity of clouds is less sensitive to
wavelength, this quadratic term may not be valid to scale the
boundaries of the HZ for planets covered by clouds''. Furthermore, the
assumption that \cite{Selsis07} make that clouds do not affect the
infrared flux from a planet at the inner edge of the habitable zone is
not strong, so the cloudy results should be interpreted with caution.
Furthermore, assumptions used in the models from which \cite{Selsis07}
derive their results assume fast planetary rotation, so it will not be
a good approximation for most tidally-locked exoplanets.

\begin{figure}
\includegraphics[width=5in]{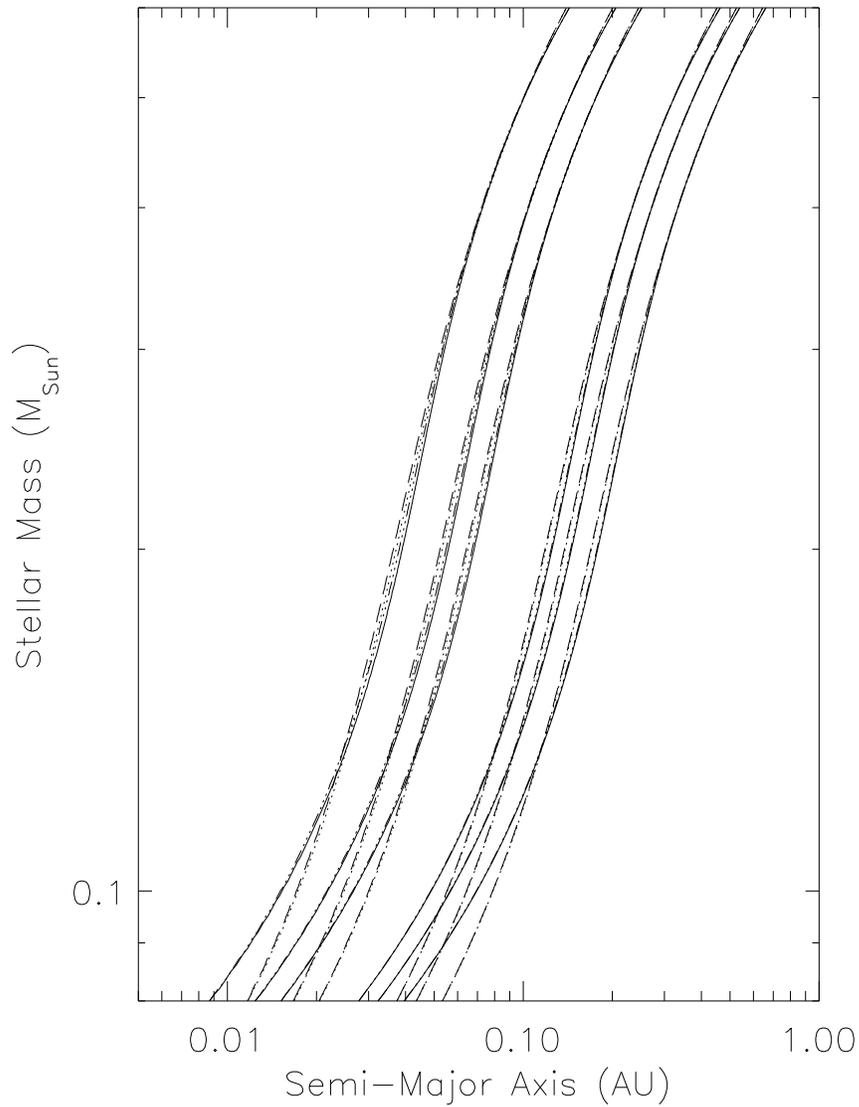}
\caption{\label{fig:ihz}IHZ boundaries for different combinations of mass-radius and
mass-luminosity relations, and using the three different cloud cover
IHZ limits from \cite{Selsis07}. Except for the inner edge near $M_* =
0.2 \msun$, the effective temperature does not affect the boundaries
(note the numerous curves that are visible). Six combinations are
plotted, but most are invisible as they are on top of each other. At
low $M_*$, the more interior IHZ limits assume the RH00
mass-luminosity relationship, see App.~\ref{app:relations}.}
\end{figure}

Fig.~\ref{fig:ihz} shows the range of the IHZ for the M dwarf mass
range for the empirical relations presented in App.~\ref{app:relations}.  For very low-mass stars ($< 0.1 \msun$), the position of the IHZ
may differ by 50\% depending on the relationships invoked. However,
for larger stars, the IHZ is mostly independent of the chosen
mass-luminosity and mass-radius relationships.

%
%
%
%

\section{The Desiccation Timescale\label{app:atmloss}}

As discussed in $\S$~\ref{sec:innerIHZ}, the details of mass loss on a
planet near the inner edge of the HZ of a very low mass star are
largely unknown. Therefore the timescale for desiccation is difficult
to constrain. In this appendix we describe a simple model for
$t_{des}$ and justify the value used in the main text of $10^8$ years.

We use the atmospheric mass loss model described in \cite{Erkaev07},
which improves the standard model by \cite{Watson81}. In the Watson et
al. picture, a layer in the atmosphere exists where absorption of high
energy photons heats the particles to a temperature that permits Jeans
escape. Photons in the X-Ray and extreme ultraviolet  have the energy to drive
this escape on most self-consistent atmospheres. In essence, the
particles carry away the excess solar energy.  

\cite{Watson81} calculated that Venus would lose its water in 280~Myr. The 
actual value is probably less than that, as they were unaware that XUV 
emission is larger for younger stars, as the spacecraft capable of such 
observations had yet to be launched. Furthermore, their model did not 
consider the possibility of mass-loss through Lagrange points, or ``Roche 
lobe overflow.'' \cite{Erkaev07} provided a slight modification the 
\cite{Watson81} model that accounts for this phenomenon for hot Jupiters. 
The physics of mass loss from a terrestrial atmosphere may be different 
than that on a giant, but as we are interested in objects that are 
experiencing tidal forces, we use the \cite{Erkaev07} model,

\begin{equation}
\label{eq:massloss}
\frac{dM_p}{dt} = - \frac{\pi R_x^2R_p\epsilon F_{XUV}}{GM_pk_{tide}},
\end{equation}
where $R_x$ is the radius of the atmosphere at which the optical depth for
stellar XUV photons is unity, $\epsilon$ is the efficiency of
converting these photons into the kinetic energy of escaping
particles, $F_{XUV}$ is the incident flux of the photons. The
parameter $k_{tide}$ represents the reduction in energy required for 
escape of a particle due to the star's gravitational pull:
\begin{equation}
\label{eq:ktide}
k_{tide} = 1 - \frac{3}{2\chi} + \frac{1}{2\chi^3} < 1,
\end{equation}
and 
\begin{equation}
\chi = \Big(\frac{M_P}{3M_*}\Big)^3\frac{a}{R_x}
\end{equation}
is the ratio of the ``Hill radius'' to the radius at the absorbing
layer. From Eq.~(\ref{eq:massloss}) it is trivial to show that 
\begin{equation}
\label{eq:tdes}
t_{des} = \frac{Gm_pm_Hk_{tide}}{\pi R_x^2R_p\epsilon F_{XUV}},
\end{equation}
where $m_H$ is the total mass of all the hydrogen atoms that must be
lost. 

The values of $F_{XUV}$ and $\epsilon$ are therefore the key
parameters as they represent the magnitude and efficiency of the
process. Unfortunately very little is known about the former for very
low-mass M dwarfs with ages of a few hundred Myr. The stars are very
faint, and moreover are prone to flaring. However, there is a secular
decrease in activity as the stars age \citep{West08}, but the data are
still too sparse to allow a satisfactory fit. Recent work connected
X-Ray sources from space telescopes (\eg \textit{GALEX}), and
ground-based surveys (\eg Sloan Digital Sky Survey), and found that
older low-mass stars tend to emit about $10^{-6}$--$10^{-3}$ of their
energy in the XUV in quiescence \citep{JonesWest12}. During flares,
that emission can increase by an order of magnitude
\citep{HawleyPettersen91,Kowalski12}. As all stars are more active when
they are young, we should expect the XUV emission to be substantially
larger at those times.

The value of $\epsilon$ must be less than 1, as not all the absorbed
energy removes particles. Observations of hot Jupiters suggest values of
$\epsilon$ of ~0.4 \citep{Yelle04,Lammer09,Jackson10}, while on Venus
$\epsilon \sim 0.15$ \citep{Chassefiere96}. However, our model also
requires these photons to dissociate the water molecules, hence we
should expect $\epsilon$ on a Tidal Venus to be much less than on a
hot Jupiter with a predominantly hydrogen atmosphere. For reference,
using the assumptions of \cite{Watson81}, we find $\epsilon = 1.7 \times
10^{-4}$ implies $t_{des} = 280$~Myr for Venus.

\begin{figure}
\includegraphics[width=5in]{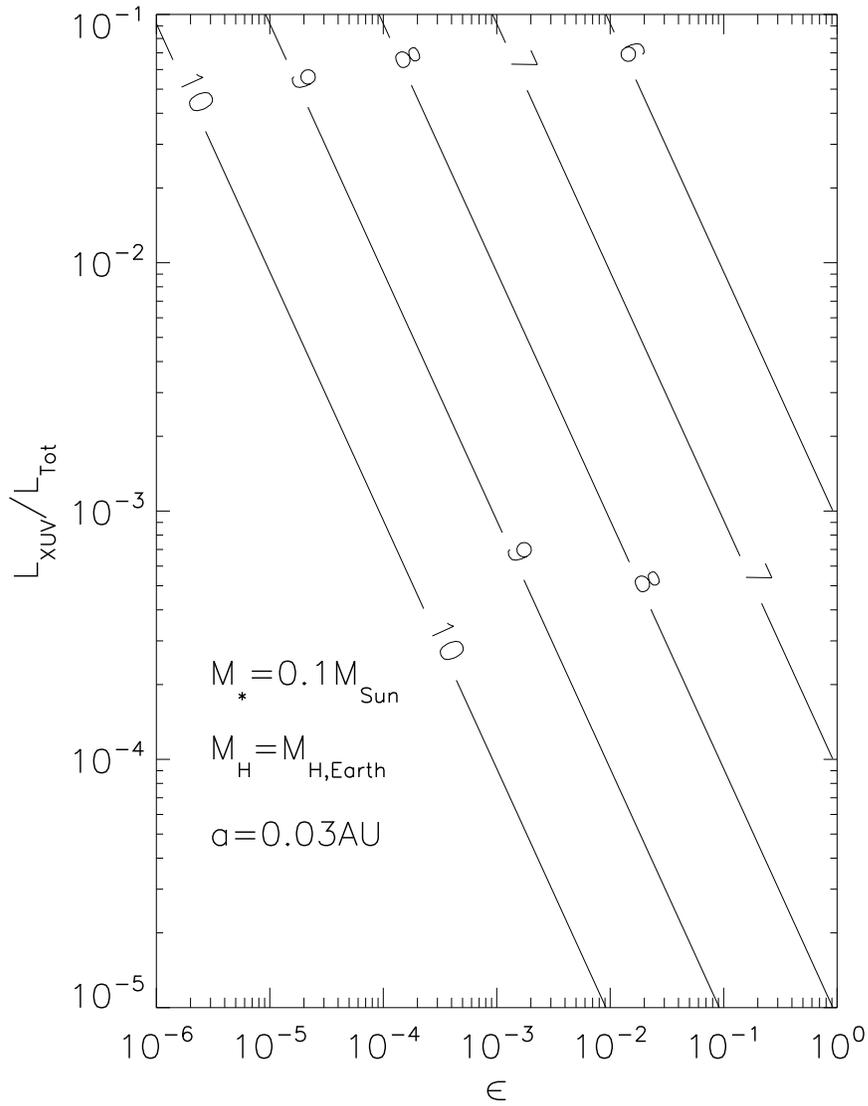}
\caption{\label{fig:tdes}Desiccation timescale for an Earth-like planet orbiting a 0.1~$\msun$ star as a function of the fraction of the luminosity in the XUV and the efficiency of converting that energy into escaping particles. Contour lines are log$_{10}(t_{des}/\textrm{yr})$. The planet initially has the same water mass fraction as the Earth and a semi-major axis of 0.03~AU.}
\end{figure}

In Fig.~\ref{fig:tdes} we show the desiccation timescale for an
Earth-like planet orbiting a 0.1~$\msun$ star at 0.03~AU for a range
of XUV luminosities and $\epsilon$. The initial mass of the planet in
hydrogen is $1.4 \times 10^{-5}~\mearth$, the current value of the
Earth and the same value used in \cite{Watson81}, and $R_x = 1.02R_p$. The luminosity of
the star is $0.0008~\lsun$, and the planet lies inside the 0\% cloud
cover HZ (see App.~\ref{app:dg}). The contour lines show
log$_{10}(t_{des})$. For $\epsilon \gsim 10^{-4}$, a few percent of the
luminosity would need to be in the XUV for $t_{des}$ to be $10^8$
years, while hot Jupiter escape rates would only require $\gsim 10^{-5}$ of
the luminosity to be in the XUV.

The environment near a forming star is very energetic and therefore we
should expect other phenomena, such as ablation and magnetic fields,
to also erode the atmosphere. Therefore the actual values of $t_{des}$
are assuredly shorter than those presented in Fig.~\ref{fig:tdes}. We
conclude that $t_{des} = 100$~Myr is a reasonable choice, while
explicitly acknowledging a large range is possible.

Finally, we note that simulations should couple the evolution of the XUV
output to the tidal orbital evolution of the planet in order to calculate
$t_{des}$ for a given system. \cite{Jackson10} created such a model
for CoRoT-7 b \citep{Leger09}, which orbits close to solar-mass
star. For such stars the generic evolution of their XUV evolution is
known \citep{Ribas05} and hence such an analysis was quite
revealing. Given the poor constraints on XUV evolution for M dwarfs, we eschewed such coupling, deferring that analysis to future work.

%
%
%
%

\section{Masses, Radii and Luminosities of M Dwarfs \label{app:relations}}

In this appendix we review relationships between mass, radius,
luminosity and effective temperature for M dwarfs ($M_* \le 0.6~\msun$), 
and their effects on
the limits of the IHZ. Measurements of these fundamental stellar
properties are challenging and several studies have produced empirical
and theoretical relations.

\begin{figure}
\includegraphics[width=5in]{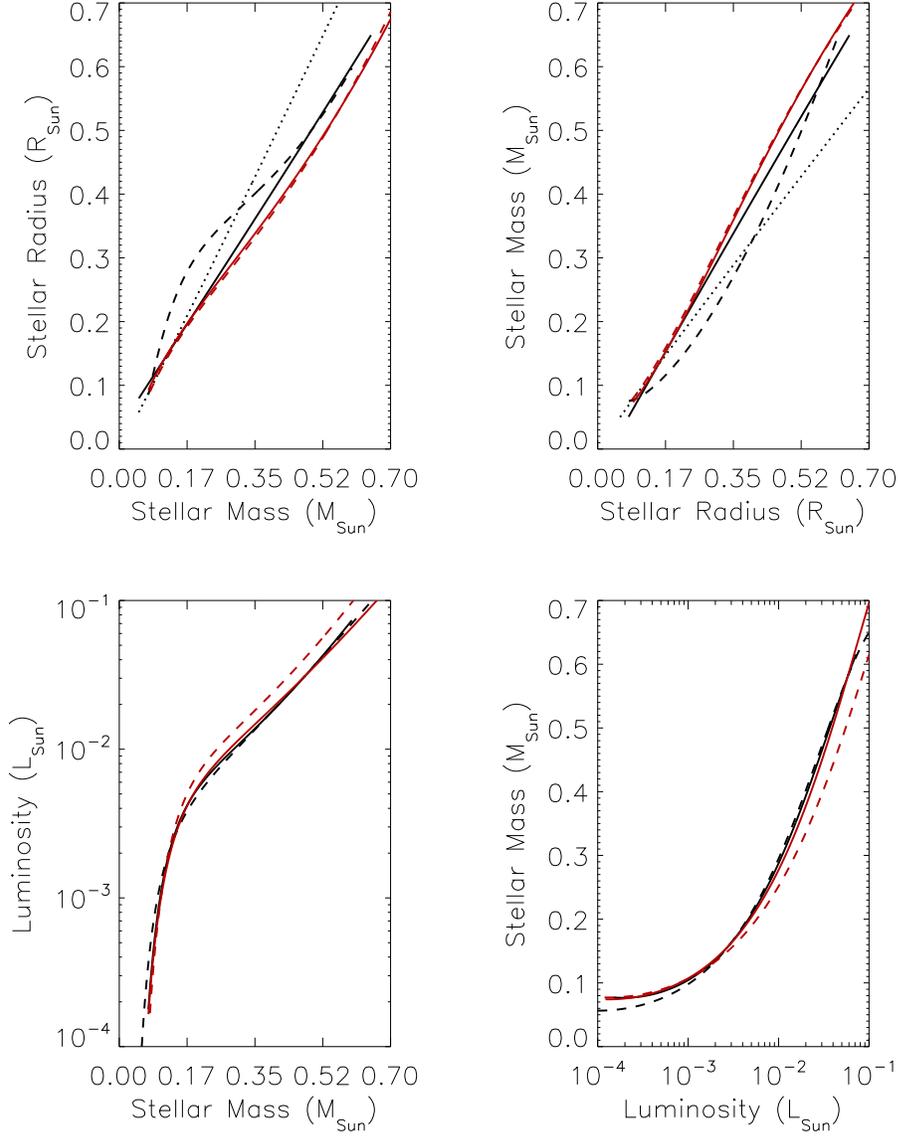}
\caption{\label{fig:relations}Scaling relations for M dwarfs. {\it Top:}
Mass-radius relations.  Black curves are empirical relations: solid from BO06, dashed from RH00, and dotted from GS99. Red curves are theoretical curves from B98: solid assumes [Fe/H] = 0, dashed [Fe/H] = -0.5. {\it Bottom:} Mass-luminosity relations.  Black curves are empirical relations: solid from RH00, and dotted from \cite{Scalo07}. Red curves are theoretical curves from B98: solid assumes [Fe/H] = 0, dashed [Fe/H] = -0.5.}
\end{figure}

We first consider the mass-radius relation. We include empirical models by 
\cite{GordaSvechnikov99} (GS99), \cite{ReidHawley00} (RH00),
\cite{BaylessOrosz06}(BO06), as well as the theoretical models of
\cite{Baraffe98} (B98). This latter model is parametrized in terms of the metallicity as defined by astronomers:
\begin{equation}\label{eq:metallicity}
[\textrm{Fe/H}] \equiv \log_{10}[\textrm{Fe/H}]_* - \log_{10}[\textrm{Fe/H}]_{Sun},
\end{equation}
in other words it is the difference between the ratio of iron to
hydrogen in a star to that of the sun. B98 consider solar
metallicity stars ([Fe/H] = 0) and sub-solar 
([Fe/H] = -0.5). GS99 and BO06 present analytic fits that are shown in
Table~1. For RH00, we take their data in Table 4.1 and derive a
third-order polynomial fit, $y = b_0 + b_1x + b_2x^2 + b_3x^3$, using
a Levenberg-Marquardt minimization scheme, and present the results in
Table~1. The top two panels of Fig.~\ref{fig:relations} show the
relationships graphically.

Next we examine the mass-luminosity relationships detailed in RH00,
\cite{Scalo07} (S07) and B98. For the former, we again fit their data with a
third-order polynomial. The fits are listed in Table~1, and shown
graphically in the bottom panels of Fig.~\ref{fig:relations}.

\clearpage
\footnotesize
\begin{center}Table 1: Scaling Relations for M Dwarfs\\
\begin{tabular}{cccccccc}
\hline\hline
$x$ & $y$ & $b_0$ & $b_1$ & $b_2$ & $b_3$ & $\chi^2$ & Ref.\\
\hline
log$_{10}$($\frac{M_*}{\msun}$) & log$_{10}$($\frac{R_*}{\rsun}$) & 0.1 & 1.03 & 0 & 0 & n/a & GS99\\
log$_{10}$($\frac{M_*}{\msun}$) & log$_{10}$($\frac{R_*}{\rsun}$) & 0.0128 & 2.185 & 3.1349 & 1.903 & 0.0136 & RH00\\
$\frac{M_*}{\msun}$ & $\frac{R_*}{\rsun}$ & 0.0324 & 0.9343 & 0.0374 & 0 & n/a & BO06\\
log$_{10}$($\frac{M_*}{\msun}$) & log$_{10}$($\frac{R_*}{\rsun}$) & 0.0372 & 1.5 & 1.09 & 0.5361 & $1.72\times 10^{-4}$ & B98([Fe/H]=0)\\
log$_{10}$($\frac{M_*}{\msun}$) & log$_{10}$($\frac{R_*}{\rsun}$) & 0.0664 & 1.685 &
1.398 & 0.7139 & $4.38\times 10^{-4}$ & B98([Fe/H]=-0.5)\\

log$_{10}$($\frac{R_*}{\rsun}$) & log$_{10}$($\frac{M_*}{\msun}$) & -0.0971 & 0.971 & $-2.501\times 10^{-5}$ & $-1.34 \times 10^{-5}$ & $1.2\times 10^{-10}$ & GS99\\
log$_{10}$($\frac{R_*}{\rsun}$) & log$_{10}$($\frac{M_*}{\msun}$) & 0.1424 & 1.568 & -0.2342 & -0.5581 & 0.0331 & RH00\\
$\frac{R_*}{\rsun}$ & $\frac{M_*}{\msun}$ & -0.03477 & 1.071 & $-8.171\times 10^{-4}$ & -0.0412 & $2.83\times 10^{-9}$ & BO06\\
log$_{10}$($\frac{R_*}{\rsun}$) & log$_{10}$($\frac{M_*}{\msun}$) & -0.03 & 0.507 & -1.156 & -0.5978 & $8.88\times 10^{-4}$ & B98([Fe/H]=0)\\
log$_{10}$($\frac{R_*}{\rsun}$) & log$_{10}$($\frac{M_*}{\msun}$) & -0.0406 & 0.4537 & -1.211 & -0.6427 & $1.39\times 10^{-3}$ & B98([Fe/H]=-0.5)\\

log$_{10}$($\frac{M_*}{\msun}$) & log$_{10}$($\frac{L_*}{\lsun}$) & 0.2984 & 8.7116 & 11.562 & 6.241 & 0.0155 & RH00\\
log$_{10}$($\frac{M_*}{\msun}$) & log$_{10}$($\frac{L_*}{\lsun}$) & 0.065 & 7.108 & 8.162 & 4.101 & n/a & S07\\
log$_{10}$($\frac{M_*}{\msun}$) & log$_{10}$($\frac{L_*}{\lsun}$) & 0.0182 & 7.228 & 9.357 & 5.281 & 0.039 & B98([Fe/H]=0)\\
log$_{10}$($\frac{M_*}{\msun}$) & log$_{10}$($\frac{L_*}{\lsun}$) & 0.4377 & 8.908 & 12.21 & 6.896 & 0.094 & B98([Fe/H]=-0.5)\\

log$_{10}$($\frac{L_*}{\lsun}$) & log$_{10}$($\frac{M_*}{\msun}$) & -0.3076 & -0.51 & -0.4504 & -0.06852 & $2.87\times 10^{-3}$ & RH00\\
log$_{10}$($\frac{L_*}{\lsun}$) & log$_{10}$($\frac{M_*}{\msun}$) & -0.3536 & -0.5463 & -0.4422 & -0.06243 & 0.224 & S07\\
log$_{10}$($\frac{L_*}{\lsun}$) & log$_{10}$($\frac{M_*}{\msun}$) & -0.04 & -0.1144 & -0.2765 & -0.045 & 0.039 & B98([Fe/H]=0)\\
log$_{10}$($\frac{L_*}{\lsun}$) & log$_{10}$($\frac{M_*}{\msun}$) & -0.03 &
$5.09\times 10^{-3}$ & -0.2135 & -0.0368 & 0.0145 & B98([Fe/H]=-0.5)\\

\end{tabular}
\end{center} 

\normalsize

%
%
%
%

\section{Radii of Terrestrial Planets\label{app:terrmr}}

In this appendix we review published studies which provide analytic formulae relating terrestrial planet mass and radius. Throughout the study we appealed to the relationship derived by \cite{Sotin07}, who considered two types of planets: Earth-like composition and ``ocean planets'' that are 50\% H$_2$O by weight. The scaling relationship for the former is
\begin{equation}
\label{eq:rterrsotin07}
\frac{R_p}{\rearth} = \left\{ \begin{array}{rl}
 \Big(\frac{M_p}{\mearth}\Big)^{0.306} &\mbox{ $10^{-2}~\mearth < M_p < \mearth$} \\
 \Big(\frac{M_p}{\mearth}\Big)^{0.274} &\mbox{ $\mearth < M_p < 10~\mearth$}
       \end{array} \right.
\end{equation}
and for the latter
\begin{equation}
\label{eq:roceansotin07}
\frac{R_p}{\rearth} = \left\{ \begin{array}{rl}
 1.258\Big(\frac{M_p}{\mearth}\Big)^{0.302} &\mbox{ $10^{-2}~\mearth < M_p < \mearth$} \\
 1.262\Big(\frac{M_p}{\mearth}\Big)^{0.275} &\mbox{ $\mearth < M_p < 10~\mearth$}
       \end{array} \right.
\end{equation}

\cite{Fortney07} parametrize the radii ($M_p > 0.01~\mearth$) in terms of the ratio of ice to rock, $f_{ice}$, or rock to iron, $f_{rock}$. They assume that these two combinations are the most likely for terrestrial exoplanets and derived the following relationships:
\begin{equation}
\label{eq:ricefortney07}
\frac{R_p}{\rearth} = (0.0912f_{ice} + 0.1603)(\log_{10}\frac{M_p}{\mearth})^2 + (0.330f_{ice}+0.7387)\log_{10}\frac{M_p}{\mearth} + 0.4639f_{ice} + 1.1193,
\end{equation}
\begin{equation}
\label{eq:rironfortney07}
\frac{R_p}{\rearth} = (0.0592f_{rock} + 0.0975)(\log_{10}\frac{M_p}{\mearth})^2 + (0.2337f_{rock}+0.4938)\log_{10}\frac{M_p}{\mearth} + 0.3102f_{rock} +0.7932.
\end{equation}
In these equation $f_{ice} = 1$ corresponds to a pure iceball,
$f_{ice} = 0$ and $f_{rock} = 1$ to a pure rock (silicate) planet, and
$f_{rock} = 0$ is a planet made of pure iron. Note that
Eqs.~(\ref{eq:ricefortney07})--(\ref{eq:rironfortney07}) are analytic
fits to complex models, and hence predict slightly different values for
$R_p$ for a planet made of rock ($f_{ice} = 0$ and $f_{rock} =
1$). The Earth has $f_{rock} = 0.67$ and hence
Eq.~(\ref{eq:rironfortney07}) becomes $R_p/\rearth =
0.1372[\log_{10}(M_p/\mearth)]^2 + 0.6504\log_{10}(M_p/\mearth) +
1.0010$.

\cite{Valencia07} considered planets in the range 1--10~$\mearth$ and parametrized the radius in terms of $f_{ice}$ \citep[see also][]{Valencia10}. They find
\begin{equation}
\label{eq:ricevalencia07}
\frac{R_p}{\rearth} = (1 + 0.56f_{ice})\frac{M_p}{\mearth}^{0.262(1-0.138f_{ice})}.
\end{equation}

\cite{Seager07} consider a wide range of planetary compositions and develop a general, but complicated, mass-radius relationship. Although they developed several formulae, here we only present one. They found that by casting $M_p$ and $R_p$ in terms of ``scaling values'' all planets follow the following relationship:
\begin{equation}
\label{rseager07}
\log_{10}r_s = k_1 + k_2\log_{10}m_s - k_2m_s^{k_3},
\end{equation}
where $k_1 = -0.20945$, $k_2 = 0.0804$, and $k_3 = 0.394$. The
parameters $m_s$ and $r_s$ are the scaled mass and radii,
respectively, and are defined as $m_s = m/m_1$ and $r_s = r/r_1$,
where the denominators are a scaling factor that depends on
composition and thermal effects. We will consider four
compositions. For pure iron planets, $m_1 = 5.8~\mearth$, and $r_1 =
2.52~\rearth$. For pure rock (perovskite) planets, $m_1 =
10.55~\mearth$, and $r_1 = 3.9~\rearth$. For pure water planets, $m_1
= 5.52~\mearth$ and $r_1 = 4.43~\rearth$. Finally, for an Earth-like
planet (30\% iron, 70\% perovskite), $m_1 = 6.41~\mearth$ and $r_1 =
2.84~\rearth$. Note that this formulation is only valid for $m_s < 4$.

The last relationship we review is that of \cite{Grasset09}, who cast
their model in terms only of $f_{ice}$. Their formulation is
complicated and rather than reproduce it here, the reader is referred
to their paper. We do not include the recent models put forth by \cite{Swift12}, as they do not include formulae, but they do find their results are consistent with those of \cite{Fortney07} and \cite{Valencia10} for rock and iron planets.

\begin{figure}
\includegraphics{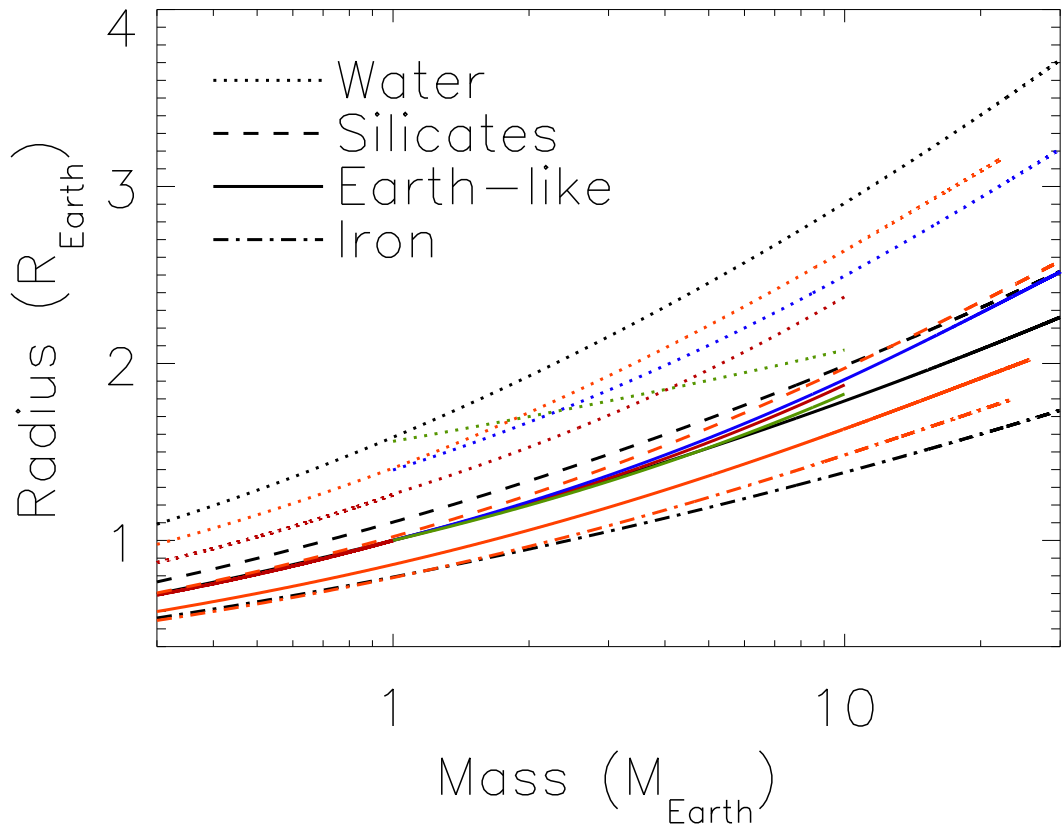}
\caption{\label{fig:terrmr}Mass-radius relationships for terrestrial planets of varying compositions. Solid curves assume a roughly Earth-like composition, dashed are all silicates, dotted are at least half water, and dot-dashed are pure iron. Black correspond to \cite{Fortney07}, red to \cite{Sotin07}, blue to \cite{Grasset09}, green to \cite{Valencia07}, and orange to \cite{Seager07}.}
\end{figure}

In Fig.~\ref{fig:terrmr} we compare these relationships from 0.3 --
30~$\mearth$, and for a variety of compositions. Not all the
relationships span that mass range, nor do they encompass all
compositions. Furthermore, although we group planets into four types,
not all types are identical. For example, we labeled the
\cite{Sotin07} ocean planet as a ``Water'' planet, but it is only 50\%
water, while the other three are 100\% water. Nonetheless, the plot
shows that if only mass is known, the radius of a planet can only be
known to a factor of 2--2.5 (of course, its terrestrial nature is also
unlikely to be known). As tidal heating scales as $R_p^5$, this
translates to a difference in heating of 30-100, assuming identical
$Q$ or $\tau$. Note that the solid red curve (Earth-like planet from
\cite{Sotin07}) has a typical shape for its class, hence the results
presented here do not vary much if a different model is chose, unless a significantly
different composition is invoked.

%
%
%
%

\section{Tidal Theory\label{app:tides}}

For our calculations of tidal evolution, we employ ``equilibrium
tide'' models, originally conceived by
\citet{Darwin1880}. This model assumes the gravitational
potential of the tide raiser can be expressed as the sum of Legendre
polynomials (\ie surface waves) and that the elongated equilibrium shape of the perturbed
body is slightly misaligned with respect to the line which connects
the two centers of mass. This misalignment is due to dissipative
processes within the deformed body and leads to a secular evolution of
the orbit as well as the spin angular momenta of the two bodies. As
described below, this approach leads to a set of 6 coupled, non-linear
differential equations, but note that the model is, in fact, linear in
the sense that there is no coupling between the surface waves which
sum to the equilibrium shape. A substantial body of research is devoted
to tidal theory
\cite[\eg][]{Hut81,FerrazMello08,Wisdom08,EfroimskyWilliams09,Leconte10}, and the reader is
referred to these studies for a more complete description of the
derivations and nuances of tidal theory. For this investigation, we
will use the models and nomenclature of \cite{Heller11}, which are
presented below.

\subsection{The Constant Phase Lag Model}

In the ``constant-phase-lag'' (CPL) model of tidal evolution, the angle between the
 line connecting the
centers of mass and the tidal bulge is assumed to be constant. This
approach has the advantage of being analogous to a damped driven
harmonic oscillator, a well-studied system, and is quite commonly
utilized in planetary studies \citep[e.g.][]{GoldreichSoter66,Greenberg09}. In this case, the
evolution is described by the following equations

\begin{equation}\label{eq:e_cpl}
  \frac{\mathrm{d}e}{\mathrm{d}t} \ = \ - \frac{ae}{8 G M_1 M_2}
  \sum\limits_{i = 1}^2Z'_i \Bigg(2\varepsilon_{0,i} - \frac{49}{2}\varepsilon_{1,i} + \frac{1}{2}\varepsilon_{2,i} + 3\varepsilon_{5,i}\Bigg)
\end{equation}

\begin{equation}\label{eq:a_cpl}
  \frac{\mathrm{d}a}{\mathrm{d}t} \ = \ \frac{a^2}{4 G M_1 M_2}
  \sum\limits_{i = 1}^2 Z'_i  \ {\Bigg(} 4\varepsilon_{0,i} + e^2{\Big [} -20\varepsilon_{0,i} + \frac{1
47}{2}\varepsilon_{1,i} + \nonumber \frac{1}{2}\varepsilon_{2,i} - 3\varepsilon_{5,i} {\Big ]} -4\sin^2(\psi_i){\Big [}\varepsilon_{0,i}-\varepsilon_{8,i}{\Big ]}{\Bigg )} 
\end{equation}

\begin{equation}\label{eq:o_cpl}
  \frac{\mathrm{d}\omega_i}{\mathrm{d}t} \ = \ - \frac{Z'_i}{8 M_i r_{\mathrm{g},i}
^2 R_i^2 n} {\Bigg (}4\varepsilon_{0,i} + e^2{\Big [} -20\varepsilon_{0,i} + 49\varepsilon_{1,i} + \varepsilon_{2,i} {\Big ]} + \nonumber \ 2\sin^2(\psi_i) {\Big [} -
2\varepsilon_{0,i} + \varepsilon_{8,i} + \varepsilon_{9,i} {\Big ]} {\Bigg )} \\  
\end{equation}

\begin{equation}\label{eq:psi_cpl}
  \frac{\mathrm{d}\psi_i}{\mathrm{d}t} \ = \ \frac{Z'_i \sin(\psi_i)}{4 M_i r_{\mathrm{g},i}^2 R_i^2 n \omega_i} {\Bigg (} {\Big [} 1-\xi_i {\Big ]}\varepsilon_{0,i} 
+ {\Big [} 1+\xi_i {\Big ]}{\Big \{}\varepsilon_{8,i}-\varepsilon_{9,i}{\Big \}} {\Bigg)} \ ,
\end{equation}

\noindent where $e$ is eccentricity, $t$ is time, $a$ is semi-major axis, $G$ is
Newton's gravitational constant, $M_1$ and $M_2$ are the two masses,
$R_1$ and $R_2$ are the two radii, $\omega_i$ are the rotational
frequencies, $\psi_i$ are the obliquities, and $n$ is the mean motion. The
quantity $Z'_i$ is

\begin{equation}\label{eq:Zp}
Z'_i \equiv 3 G^2 k_{2,i} M_j^2 (M_i+M_j) \frac{R_i^5}{a^9} \ \frac{1}{n Q_i} \ ,
\end{equation}

\noindent where $k_{2,i}$ are the Love numbers of order 2, and $Q_i$ are the ``tidal quality factors.'' The parameter $\xi_i$ is

\begin{equation}\label{eq:chi}
\xi_i \equiv \frac{r_{\mathrm{g},i}^2 R_i^2 \omega_i a n }{ G M_j},
\end{equation}

\noindent where $i$ and $j$ refer to the two bodies, and $r_g$ is the ``radius of gyration,'' \ie the moment of inertia is $M(r_gR)^2$. The signs of the phase lags are

\begin{equation}\label{eq:epsilon}
\begin{array}{l}
\varepsilon_{0,i} = \Sigma(2 \omega_i - 2 n)\\
\varepsilon_{1,i} = \Sigma(2 \omega_i - 3 n)\\
\varepsilon_{2,i} = \Sigma(2 \omega_i - n)\\
\varepsilon_{5,i} = \Sigma(n)\\
\varepsilon_{8,i} = \Sigma(\omega_i - 2 n)\\
\varepsilon_{9,i} = \Sigma(\omega_i) \ ,\\
\end{array}
\end{equation}

\noindent with $\Sigma(x)$ the sign of any physical quantity $x$, thus
$\Sigma(x)~=~+~1~\vee~-~1~\vee~0$. 

The tidal heating of the $i$th body is due to the transformation of
rotational and/or orbital energy into frictional heating. The heating
from the orbit is 

\begin{equation}\label{eq:E_orb_cpl}
\nonumber
\dot{E}_{\mathrm{orb},i} = \ \frac{Z'_i}{8} \ \times \ {\Big (} \ 4 \varepsilon_{0,i} + e^2 {\Big [-20 \varepsilon_{0,i} + \frac{147}{2} \varepsilon_{1,i} + \frac{1}{2} \varepsilon_{2,i} - 3 \varepsilon_{5,i} {\Big ]} - 4 \sin^2(\psi_i) \ {\Big [}\varepsilon_{0,i} - \varepsilon_{8,i}{\Big ]} \ {\Big )}} \ ,
\end{equation}

\noindent and that from the rotation is 

\begin{equation}\label{eq:E_rot_cpl}
\nonumber
\dot{E}_{\mathrm{rot},i} = \ - \frac{Z'_i}{8} \frac{\omega_i}{n} \ \times \ {\Big (} \ 4 \varepsilon_{0,i} + e^2 {\Big [}-20 \varepsilon_{0,i} + 49 \varepsilon_{1,i} 
+ \varepsilon_{2,i}{\Big ]} + 2 \sin^2(\psi_i) \ {\Big [}- 2 \varepsilon_{0,i} + \varepsilon_{8,i} + \varepsilon_{9,i}{\Big ]} 
\ {\Big )} \ .
\end{equation}

\noindent The total heat in the $i$th body is therefore 

\begin{equation}\label{eq:E_tide_cpl}
\dot{E}_{\mathrm{tide},i}^{\mathrm{CPL}} = - \ (\dot{E}_{\mathrm{orb},i} + \dot{E}_{\mathrm{rot},i}) > 0 \ .
\end{equation}
The rate of evolution and amount of heating are set by three free
parameters: $Q, k _2$, and $r_g$. We discuss the ranges of these
parameters below in App.~\ref{app:tides}.4. 

\begin{figure}
\includegraphics{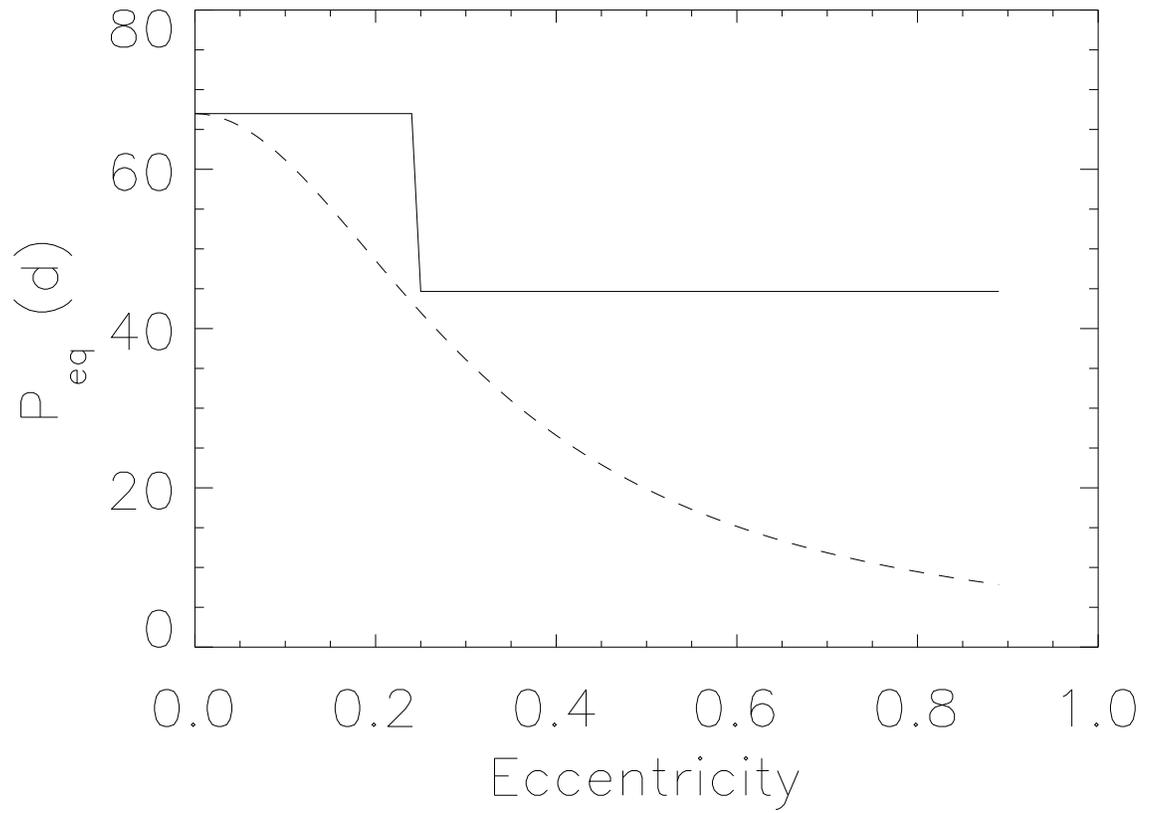}
\caption{\label{fig:eqspin}Comparison of the equilibrium spin periods in
the CPL framework. The solid line is the value predicted by our model,
Eqs.~(\ref{eq:e_cpl})--(\ref{eq:epsilon}), the dashed is the
``continuous'' model of \cite{Goldreich66}.}
\end{figure}

\cite{Goldreich66} suggested that the equilibrium rotation period for both bodies is

\begin{equation}\label{eq:p_eq_cpl}
P_{eq}^{CPL} = \frac{P}{1 + 9.5e^2}.
\end{equation}
\cite{MurrayDermott99} present a derivation of this expression, which
assumes the rotation rate may take a continuum of values. However, the
CPL model described above only permits 4 ``tidal waves'', and hence does
not permit this continuum. In Fig.~\ref{fig:eqspin} we compare
\eqref{eq:p_eq_cpl} to the equilibrium values predicted by
Eqs.~(\ref{eq:e_cpl})--(\ref{eq:epsilon}) as a function of
eccentricity for Gl 581 d, \ie an orbital period of 66 days, a mass of
$5.6~\mearth$ and a radius of $1.6~\rearth$
\citep{Mayor09}. As expected \eqref{eq:p_eq_cpl} predicts a continuous
range of periods, whereas our model (solid curve) does not. The jump
at $e = \sqrt{1/19}$ occurs at the $\omega/n = 1.5$ transition, and
nearly reaches the value predicted by \eqref{eq:p_eq_cpl}. The next phase
jump at $e = \sqrt{2/19} \approx 0.32$ is not included in the CPL model, and hence it breaks down at larger values. However, at larger $e$ values,
coupling between waves and other non-linear effects may dampen the
role of tides. Therefore the evolution at larger $e$ predicted by the
CPL model may not be qualitatively correct. We urge caution when
interpreting CPL results above $e = 0.32$.

Thus, the CPL model can be considered from two perspectives. When
modeling the evolution of a system, one should use the discrete spin
values for self-consistency, \ie as an initial condition, or if forcing the spin to remain tide-locked. However, if calculating the equilibrium
spin period separately, the continuous value of Eq.~\ref{eq:p_eq_cpl}
should be used. We refer to these rotations as ``discrete states'' and
the ``continuous state''. Note that this point marks a difference
between our implementation of the CPL model, and that described in
\cite{Heller11}.

\subsection{The Constant Time Lag Model}

The constant-time-lag (CTL) model assumes that the time interval
between the passage of the perturber and the tidal bulge is
constant. This assumption allows the tidal response to be continuous
over a wide range of frequencies, unlike the CPL model. But, if
the phase lag is a function of the forcing frequency, then the system is no longer analogous to a
damped driven harmonic oscillator. Therefore, this model should only
be used over a narrow range of frequencies, see
\cite{Greenberg09}. However, this model's use is widespread,
especially at high $e$, so we use it to evaluate tidal effects as
well. This model predicts larger tidal heating and evolution rates
at high $e$ as any coupling between tidal waves is ignored. Therefore,
the CPL and CTL models probably bracket the actual evolution.

The evolution is described by the following equations:

\begin{equation} \label{eq:e_ctl}
  \frac{\mathrm{d}e}{\mathrm{d}t} \ = \ \frac{11 ae}{2 G M_1 M_2}
  \sum\limits_{i = 1}^2Z_i \Bigg(\cos(\psi_i) \frac{f_4(e)}{\beta^{10}(e)}  \frac{\omega_i}{n} -\frac{18}{11} \frac{f_3(e)}{\beta^{13}(e)}\Bigg)
\end{equation}

\begin{equation}\label{eq:a_ctl}
  \frac{\mathrm{d}a}{\mathrm{d}t} \ = \  \frac{2 a^2}{G M_1 M_2}
  \sum\limits_{i = 1}^2 Z_i \Bigg(\cos(\psi_i) \frac{f_2(e)}{\beta^{12}(e)} \frac{\omega_i}{n} - \frac{f_1(e)}{\beta^{15}(e)}\Bigg)
\end{equation}

\begin{equation}\label{eq:o_ctl}
  \frac{\mathrm{d}\omega_i}{\mathrm{d}t} \ = \ \frac{Z_i}{2 M_i r_{\mathrm{g},i}^2 
R_i^2 n} \Bigg( 2 \cos(\psi_i) \frac{f_2(e)}{\beta^{12}(e)} - \left[ 1+\cos^2(\psi)
 \right] \frac{f_5(e)}{\beta^9(e)} 
\frac{\omega_i}{n} \Bigg)  
\end{equation}

\begin{equation}\label{eq:psi_ctl}
  \frac{\mathrm{d}\psi_i}{\mathrm{d}t} \ = \ \frac{Z_i \sin(\psi_i)}{2 M_i r_{\mathrm{g},i}^2 R_i^2 n \omega_i}\left( \left[ \cos(\psi_i) - \frac{\xi_i}{ \beta} \right] \frac{f_5(e)}{\beta^9(e)} \frac{\omega_i}{n} - 2 \frac{f_2(e)}{\beta^{12}(e)} \right)
\end{equation}

\noindent where

\begin{equation}\label{eq:Z}
 Z_i \equiv 3 G^2 k_{2,i} M_j^2 (M_i+M_j) \frac{R_i^5}{a^9} \ \tau_i \ ,
\end{equation}

\noindent and 

\begin{equation}\label{eq:f_e}
\begin{array}{l}
\beta(e) = \sqrt{1-e^2},\\
f_1(e) = 1 + \frac{31}{2} e^2 + \frac{255}{8} e^4 + \frac{185}{16} e^6 + \frac{25}{
64} e^8,\\
f_2(e) = 1 + \frac{15}{2} e^2 + \frac{45}{8} e^4 \ \ + \frac{5}{16} e^6,\\
f_3(e) = 1 + \frac{15}{4} e^2 + \frac{15}{8} e^4 \ \ + \frac{5}{64} e^6,\\
f_4(e) = 1 + \frac{3}{2} e^2 \ \ + \frac{1}{8} e^4,\\
f_5(e) = 1 + 3 e^2 \ \ \ + \frac{3}{8} e^4.
\end{array}
\end{equation}

\noindent The tidal heating of the $i$th body is therefore

\begin{equation}\label{eq:E_tide_ctl}
\dot{E}_{\mathrm{tide},i}^{\mathrm{CTL}} = \ Z_i {\Bigg (} \frac{f_1(e)}{\beta^{15}
(e)} - 2 \frac{f_2(e)}{\beta^{12}(e)} \cos(\psi_i) \frac{\omega_i}{n} + \ {\Big [} 
\frac{1+\cos^2(\psi_i)}{2} {\Big ]} \frac{f_5(e)}{\beta^9(e)}\Big({\frac{\omega_i}{n}}\Big)^2 {\Bigg )} \ .
\end{equation}
Note that Eqs.~(\ref{eq:E_tide_cpl}) and (\ref{eq:E_tide_ctl}) include
terms due to both rotation and obliquity, which were not discussed in
the main text. Should information regarding these parameters become
available for particular planets, they should be included in estimates
of tidal heating and habitability.

It can also be shown that the equilibrium rotation period for both bodies is
\begin{equation}\label{eq:p_eq_ctl_obl}
P_{eq}^{CTL}(e,\psi) = P\frac{\beta^3f_5(e)(1 + \cos^2\psi)}{2f_2(e)\cos\psi},
\end{equation}
which for low $e$ and $\psi = 0$ reduces to
\begin{equation}\label{eq:p_eq_ctl}
P_{eq}^{CTL} = \frac{P}{1 + 6e^2}.
\end{equation}

There is no general conversion between $ Q_\mathrm{p}$ and
$\tau_\mathrm{p}$. Only if $e~=~0$ and $\psi_\mathrm{p}~=~0$, when
merely a single tidal lag angle $\varepsilon_\mathrm{p}$ exists, then
\begin{equation}\label{eq:qtau}
Q_\mathrm{p}~\approx~1/(2|n-\omega_\mathrm{p}|\tau_\mathrm{p}),
\end{equation}
as long as $n-\omega_\mathrm{p}$ remains unchanged. Hence,
a dissipation value for an Earth-like planet of $Q_\mathrm{p}~=~100$
is not necessarily equivalent to a tidal time lag of 638\,s, so the results for
the tidal evolution will intrinsically differ among the CPL and the
CTL model. However, both choices are common for the respective model.

\subsection{Numerical Methods}

The rates of change of the 6 parameters affected by tidal theory
are all different. The orbital parameters tend to evolve several
orders of magnitude more slowly than the planetary spin parameters, at
least for the cases we are interested in. Therefore we implement a
dynamical timestep scheme in our integration of tidal evolution. At
the beginning of each timestep, we calculate the derivatives and set
the timestep to
\begin{equation}
\label{eq:dt}
\Delta t =
\eta \times \mathrm{min}\Big(\frac{a}{da/dt},\frac{e}{de/dt},\frac{\omega_*}{d\omega_*/dt},\frac{\psi_*}{d\psi_*/dt},\frac{\omega_p}{d\omega_p/dt},\frac{\psi_p}{d\psi_p/dt}\Big),
\end{equation}
where $\eta$ is a constant, which we set to 0.01. We find that larger
values of $\eta$, even up to unity, produce results similar with $\eta
= 0.01$, but at our more conservative choice, we find almost no
difference with even smaller value of $\eta$, \ie the solution has
converged. This implementation increases the speed of the code by
several orders of magnitude in some cases. For example, if considering
a planet with spin properties similar to the Earth at 0.01 AU, the
timescale to damp out the obliquity is of order years, but the orbit
evolves in millions of years (depending on $e$). Integrating this
system for billions of years with a fixed timestep chosen to resolve
the obliquity evolution will take many orders of magnitude more time
than using this scheme. Dynamical timestepping is especially important
for planets that merge with their host star \citep{Jackson09,Levrard09}.

In many cases, $e$ and $\psi$ can damp to arbitrarily small values,
and hence d$e$/d$t$ and d$\psi$/d$t$ take on meaningless values. In
some cases, their respective timescales as defined above can become
very short, even though they are effectively 0. This situation
destroys the speed advantage gained by utilizing
\eqref{eq:dt}. We therefore set a floor for these values at
$10^{-10}$. When they reach this level, we set them to 0, and do not
include them in the determination of $\Delta t$.

Finally, care must be taken as a planet approaches tidal locking. In
many cases an integration will bounce back and forth above and below
the equilibrium period. This possibility is a
natural result of the discrete nature of numerical integration, but,
again, can diminish the benefits of a dynamical
timestep. Therefore, if a planet comes within 10\% of the equilibrium
period, we set the spin period to the equilibrium value. If the orbit
is eccentric and evolving, after tidal locking, we force the planet to
remain locked, \ie the spin period obeys
Eqs.~(\ref{eq:p_eq_cpl}) or (\ref{eq:p_eq_ctl_obl}).

\subsection{Tidal Response in Celestial Bodies}

Exoplanets are expected to have an even wider diversity of
compositions and structures than those in our Solar System
\citep[see][]{Raymond04,Leger04,OBrien06,Bond10}, thus their response to
tidal processes probably varies greatly. We therefore choose to keep
our model simple and set $k_2 = r_g = 0.5$ for all planets.

The Earth is the only body in the Solar System with a significant
amount of water on its surface, and is also the planet with the most
detailed picture of its tidal processes. The majority of tidal
dissipation on the Earth occurs in the oceans, and we assume the same
will occur on terrestrial exoplanets. Tidal dissipation in the Earth's
ocean was directly measured from {\it TOPEX} and {\it Poseiden}
satellite data by \cite{EgbertRay00} who found that 50--75\% of
dissipation occurs in and around shallow seas, with the remainder in
the open ocean, presumably due to turbulence caused by
topography. Recent computer simulations have successfully reproduced
this observation
\citep{Jayne01,Nycander07,Kelly10}. For a more complete treatment of
this process, consult \cite{Garret07}. Therefore dissipation in planetary oceans is a complex function of continental shapes, sea floor topography and the forcing frequency.

Lunar Laser Ranging, in which the timing of laser pulses bounced off
the Moon is measured, has provided very precise measurements of the
rate of the Moon's recession from the Earth \citep{Dickey94}. If this
recession is due only to tidal dissipation within the Earth, than the
Earth's tidal quality factor is $Q_{Earth} = 12$ \citep{Yoder95}. This
value is lower than most measurements in the Solar System, which are
more uncertain, see below. However, given the presence of surface
oceans, this extra dissipation should be expected. On the other hand,
extrapolating backward from the current orbital expansion rate and
employing the CPL model, one finds the Moon was at the Earth's surface
$\sim 2$~Gyr ago \citep{MacDonald64}. The moon-forming impact has been
dated to about 4.5~Gyr ago, and hence the Earth's current $Q$ appears
too low. This discrepancy may be explained by 4 possibilities: 1) The
current $Q$ is anomalously low, and in the past different continental
configurations and/or ocean floor topography did not permit such a
high dissipation, 2) The tidal effects are a function of frequency,
and as the Earth's rotation has slowed, the dissipation rate has
increased, 3) Perturbations from other planets affected the evolution
\citep{Cuk07}, or 4) Linear theory is oversimplified. Probably all
of these possibilities play a role, and make the extension of
tidal theory to an Earth-like exoplanet unreliable.

Measurements of the Martian dissipation
function find $Q_\mathrm{Mars}~=~85.58 \pm 0.37$ \citep{Bills05} or $Q_\mathrm{Mars}~=~79.91\pm0.69$ \citep[][]{Lainey07}. The value for Mercury is
$Q_\mathrm{Mercury}~<~190$, and for Venus it is $Q_\mathrm{Venus}~<~17$
\citep{GoldreichSoter66}. Observations of Io imply $Q_\mathrm{Io}~<~100$ \citep{Peale79,Segatz88}, while the Moon lies at $Q_\mathrm{Moon}~=~26.5\pm1$
\citep{Dickey94}. These measurements indicate a similar order of magnitude for
the tidal dissipation function of all desiccated bodies in the Solar System. 

Tidal effects on Venus may be particularly complicated due to its
large atmosphere. The rotation rate of Venus is such that it always
presents the same face to the Earth at inferior conjunction. Although
this may be by chance due to Venus approaching tidal lock with the
Sun, it has also been suggested that at some point in the past thermal
atmospheric tides could have provided an equal but opposite torque
than that from the rocky component. If that occurred, the torque from
the Earth could dominate and spin-lock Venus
\citep{GoldSoter69}. More recent and detailed work has shown that this
interaction can be significant and permits chaotic spin evolution that
eventually settles into 1 of 4 possible states
\citep{CorreiaLaskar01}. Although a large atmosphere can play a
significant role in the evolution of terrestrial planets
\citep{CorreiaLaskar03}, we ignore their effect here due to the large
number of unknowns for an exoplanet.

In real bodies, $Q$ is a function of the bodies' rigidity $\mu$,
viscosity $\eta$, and temperature $T$
\citep{Segatz88,Fischer90}. A comprehensive
tidal model would couple the orbital and structural evolution of the
bodies since small perturbations in $T$ can result in large variations
in $Q$
\citep{Segatz88,MardlingLin02,EfroimskyLainey07}. Moreover,
tidally-generated heat will not necessarily be transported to the
surface quickly. The internal structure of the planet may trap the
energy for long periods of time, resulting in catastrophic tectonics
when the heat is finally released, as may have occurred on Venus
\citep{Turcotte96}. This episodic volcanism could change $t_{des}$
dramatically, but, on the other hand, complete lithospheric overturn
is probably just as detrimental to life as desiccation. Nonetheless,
future work should explore these details. \citet{Henning09} performed
detailed calculations and argue that $k_2=0.3$ and $Q~=~50$ are 
reasonable choices for dry planets.

The tidal properties of stars are about as poorly constrained as for
planets. The primary constraint comes from the distribution of
orbits of binary stars \citep[\eg][]{Zahn77}, whose orbits are
circularized by the tidal interaction. However, for the closest
binaries, where the tidal effects are strongest, and one might think
the constraints the tightest, the situation is complicated by the
early phases of stellar evolution \citep{ZahnBouchet89}, in which
stellar radii contract rapidly. As the radius enters the tidal
evolution at the fifth power, the radial contraction is a major
effect. Recently \citet{KhaliullinKhaliullina11} revisited this point
and found that, although radial contraction does play a major role,
they were unable to combine models of the equilibrium tide and radial
contraction to match observations. \cite{Weinberg11} tested the
linear theory against non-linear models and found that in
general the linear theory does not capture the physics of the more
sophisticated approach. These studies demonstrate that there still
exist serious problem with the linear theory, and quantitative
discrepancies exist. However, it does qualitatively model the relevant
physics, and allows wide sweeps of parameter space to be a
computationally tractable problem. We therefore continue to use it,
but explicitly acknowledge that known weaknesses persist.

A wide range of values for the tidal quality factor and tidal time lag
for stars are present in the literature, ranging from $Q_* = 10^4$
\citep{Adams11} all the way to $Q_* = 10^9$ \citep{Matsumura10}, with
many others finding intermediate values
\citep[\eg][]{Lin96,CaronePatzold07,Jackson08,Jackson09,IbguiBurrows09}. For
the star we take an intermediate and typical value of $Q_* = 10^6$.

\end{document}